\documentclass[a4paper]{article}
\usepackage{Odyssey2020}
\usepackage{epsfig,amssymb,amsmath}
\ninept

\usepackage{color}
\usepackage{dirtytalk}
\usepackage{url}
\usepackage{mathtools}
\usepackage{subfigure}
\usepackage{multirow}
\usepackage{verbatim}
\usepackage[ruled,vlined]{algorithm2e}

\newcommand{\tablesqueezevspace}{\vspace{-.25cm}}
\newcommand{\figsqueezevspace}{\vspace{-.4cm}}
\newcommand{\sectionpresqueeze}{\vspace{-.25cm}}
\newcommand{\sectionpostsqueeze}{\vspace{-.1cm}}

\title{An initial investigation on optimizing tandem speaker verification and countermeasure systems using reinforcement learning}

\makeatletter
\def\name#1{\gdef\@name{#1\\}}
\makeatother
\name{{\em Anssi Kanervisto\textsuperscript{1},
       Ville Hautam{\"a}ki\textsuperscript{1},
       Tomi Kinnunen\textsuperscript{1},
       Junichi Yamagishi\textsuperscript{2}}
       \thanks{This work was partially funded by the Academy of Finland (projects no. 313970 and 309629) and JST CREST Grant (JPMJCR18A6, VoicePersonae project, Japan), with computing resources from IT Center of Sciences (CSC). This research was carried out when the first author was at the National Institute of Informatics (NII) of Japan in 2019. We thank Cheng-I Lai for providing code used as a reference in this work.}
}

\address{\textsuperscript{1}University of Eastern Finland, Finland.  \\
\textsuperscript{2}National Institute of Informatics, Japan. \\
{\small \tt \{anssk, villeh, tkinnu\}@cs.uef.fi,  jyamagis@nii.ac.jp}}
\begin{document}
\maketitle

\begin{abstract}
The spoofing countermeasure (CM) systems in automatic speaker verification (ASV) are not typically used in isolation of each other. These systems can be combined, for example, into a cascaded system where CM produces first a decision whether the input is synthetic or bona fide speech. In case the CM decides it is a bona fide sample, then the ASV system will consider it for speaker verification. End users of the system are not interested in the performance of the individual sub-modules, but instead are interested in the performance of the combined system. Such combination can be evaluated with tandem detection cost function (t-DCF) measure, yet the individual components are trained separately from each other using their own performance metrics. In this work we study training the ASV and CM components together for a better t-DCF measure by using reinforcement learning. We demonstrate that such training procedure indeed is able to improve the performance of the combined system, and does so with more reliable results than with the standard supervised learning techniques we compare against.
\end{abstract}

\sectionpresqueeze
\section{Introduction}
\sectionpostsqueeze
    \textit{Automatic speaker verification} (ASV) systems attempt to verify if a given speech utterance matches the claimed identity \cite{Reynolds95_speaker}. Alone such systems are susceptible to malicious attacks via replay attack \cite{kinnunen2017asvspoof} or speech synthesis \cite{todisco2019asvspoof}, where an attacker attempts to fool the ASV system with crafted audio samples. To combat this, \textit{spoofing countermeasure} (CM) systems aim to detect these malicious samples from genuine, \say{bona fide} samples. These systems can then be used along with ASV systems to detect and discard access attempts with \textit{e.g.} synthetic speech, to improve the security of the system \cite{todisco2019asvspoof}.
    
    The ASV and CM systems are trained in isolation from each other, and later combined by running test trials through both systems. To measure performance of such a combined system, the authors of \cite{kinnunen2018t} proposed \textit{tandem detection cost function} (t-DCF) to evaluate the performance of the combined system performance. This performance metric takes into account different error situations that arise from having two different systems and assigns different costs for these errors. It is a generalization of the widely adopted \textit{detection cost function} (DCF) \cite{brummer2006application} involving an extra, \say{spoof} class and two (rather than one) systems being evaluated.
    
    The original ASV and CM systems are not trained to minimize the t-DCF metric: they are trained to minimize the errors in their respective tasks, completely oblivious of the existence of the other systems and the final use-case. Our aim is to improve the metric of the final task, namely the t-DCF. Unfortunately, to compute t-DCF, one has to compute error-rates with hard decisions, which is a non-differentiable operation, preventing updating model parameters with gradient descent. One approach for optimizing count-based metrics relies on \say{softening} the counts with soft (differentiable) versions, \textit{e.g.} by methods based on \textit{maximal figure-of-merit} \cite{kukanov2016deep, gao2006maximal} or using \textit{hinge-loss} to approximate error counts \cite{sizov2017direct, bishop2006pattern}. While shown to be effective optimization techniques, these methods involve functions that tend to require tuning of parameters for efficient learning \cite{gao2006maximal}.
    
    Meanwhile, \textit{policy gradients} \cite{sutton2000policy} and \textit{REINFORCE algorithm} \cite{williams1992simple} offer a generic framework for optimizing functions in an end-to-end fashion by \textit{sampling the function}. They are the basis of reinforcement learning methods (\textit{e.g.} \cite{ppo, gae}), which provide state-of-the-art performance. Policy gradient -based methods have also been applied to situations where loss functions depend on sampled variables \cite{yu2017seqgan, jang2017categorical}. REINFORCE provides provably zero-bias estimate of the gradient of the objective function and does not have parameters to tune but suffers from the high variance in the estimated gradients \cite{jang2017categorical}.

    We propose to optimize tandem system with ASV and CM systems to perform secure speaker verification using the REINFORCE algorithm. We pre-train disjoint ASV and CM systems which are then fine-tuned for the spoofing-robust speaker verification task with REINFORCE using rewards derived from the t-DCF cost parameters. We then compare these against standard supervised approaches for fine-tuning such systems. Our novelty is two-fold. First, the use of REINFORCE on a system of multiple individual parts (ASV and CM) allows us to analyze how individual components change during training. Second, this is the first attempt of optimizing t-DCF directly for spoofing-robust speaker verification system.

\sectionpresqueeze
\section{Tandem Detection Cost Functions}
\sectionpostsqueeze
\label{sec:tdcf}
    An automatic speaker verification system compares enrollment and test utterances in terms of their speaker similarity and takes a hard \texttt{ACCEPT} (same speaker) or \texttt{REJECT} (different speaker) action. As a binary any classifier, ASV systems are prone to two different types of errors, \emph{false alarms} (false accepts) and \emph{misses} (false rejections). 
    Since the relative severity of each error depends on the application, the performance of an ASV system is often gauged using the so-called \emph{detection cost function} (DCF) which weights the two errors as \cite{brummer2006application},
        \begin{equation}
            \begin{split}
            \label{eq:basic-dcf}
            \text{DCF} & = \rho_\text{tar}C_\text{miss}P_\text{miss} + \rho_\text{non}C_\text{fa}P_\text{fa}\\
            & = \rho_\text{tar}C_\text{miss}P_\text{miss} + (1-\rho_\text{tar})C_\text{fa}P_\text{fa}\\
            \end{split}
        \end{equation}
    where $\rho_\text{tar},\rho_\text{non}$ are the \emph{prior probabilities} of genuine (target) and impostor (nontarget) users; the second line is obtained by assuming that there are only target and nontarget classes ($\rho_\text{tar} + \rho_\text{non} = 1$). Further, $C_\text{miss}, C_\text{fa}>0$ are the \emph{costs} of a miss and a false alarm, and $P_\text{miss},P_\text{fa}$ are the two \emph{error rates} of the evaluated system, obtained by hard-counting of errors. The cost parameters $(C_\text{miss}, C_\text{fa}, \rho_\text{tar})$ are set in advance by the evaluator and remain fixed in a given evaluation set-up. 
    
    While \eqref{eq:basic-dcf} is well suited for performance assessment of ASV systems used by humans only, it does not factor in the impact of \textit{spoofing attacks}. As ASV systems are easily fooled by spoofing attacks, CMs are necessary in retaining the trustworthiness of ASV systems. Here, CM is defined as another binary classifier that aims at discriminating bona fide (human) and spoofed speech utterances. A typical design is to place a CM in front of an ASV system as a human/spoof gate. While conventional ASV systems have two possible errors, the cascaded system may now experience four different types of errors: (a) a target user is accepted by the CM but rejected by the ASV system; (b) a nontarget user gets accepted by both systems; (c) a spoofing attack gets accepted by both systems; (d) a target user is rejected by the CM. With this motivation, the authors of \cite{kinnunen2018t} extended the DCF to a \emph{tandem} DCF (t-DCF), defined as\footnote{The definition \eqref{eq:new_tdcf} differs slightly from \cite{kinnunen2018t} in terms of how the cost parameters are defined but the other ingredients remain the same (related manuscript under preparation).},
            \begin{equation}
            \begin{split}
            \label{eq:new_tdcf}
            \text{t-DCF} &= C_\text{miss} \cdot \rho_\text{tar} (P_\text{a} + P_\text{d}) \\
                &+ C_\text{fa} \cdot \rho_\text{non} \cdot P_\text{b} \\
                &+ C_\text{fa,spoof} \cdot \rho_\text{spoof} \cdot P_\text{c} .
            \end{split}
        \end{equation}
    where $P_\bullet$ denote the error rates of the four cases noted above (obtained by treating the CM and ASV decisions independent), $\rho_\text{spoof}$ is the prior probability of a spoofing attack and $C_\text{fa,spoof}$ is the cost of falsely accepting a spoofing attack.
    
    
    Both \eqref{eq:basic-dcf} and \eqref{eq:new_tdcf} have primarily been cast for the purposes of \emph{evaluation} but not \emph{optimization}. All the involved error rates needed for computing these metrics are obtained by counting errors, leading to generally non-differentiable optimization problem. We could design a \say{soft t-DCF} using approximations for the hard error-counts \cite{kukanov2016deep, sizov2017direct, gao2006maximal}, but this would require selecting and tuning the parameters of the approximations or require complex derivations. Instead, we opt for reinforcement learning methods, as they offer a simple, plug-and-play algorithm to this problem which we can apply almost directly on this tandem system to train for better t-DCF.

\sectionpresqueeze
\section{Reinforcement learning}
\sectionpostsqueeze
    In this section we will cover the terminology and base concepts required for the REINFORCE algorithm, the core of the tandem optimization studied in this work.
    
    \subsection{Background on reinforcement learning}
        \textit{Reinforcement learning} aims to solve problems formulated as \textit{Markov decision processes} (MDPs) \cite{sutton2018reinforcement}. A MDP consists of five distinct components at different timesteps $t \in \mathbb{N}$: \textbf{states} $s_t \in S$, available \textbf{actions} $a_t \in A(s)$, \textbf{transition dynamics} $s_{t+1} \sim p(s_{t+1} | s_t, a_t)$, \textbf{rewards} $r_t = R(s_t, a_t, s_{t+1})$ and finally a \textbf{policy} which provides a distribution over actions $\pi(a_t | s_t)$. The MDP begins in some state $s_0 \sim p_0(s)$, followed by sampling an action $a_0 \sim \pi(a_0 | s_0)$ which is then used to evolve the MDP by one timestep $s_1 \sim p(s_1 | s_t, a_t)$. In the process MDP provides reward $r_0 = R(s_0, a_0, s_1)$. This process continues until a terminal state $s_T$ is reached, where $T \in \mathbb{N}$ denotes length of the episode. One game from beginning to the terminal state is called an \textbf{episode}, and the experienced sequence $(s_0, a_0, r_0, s_1, a_1, r_1, s_2 \ldots s_T)$ is called a \textbf{trajectory}. It is common to assume that the transition dynamics is not available. We do not know, for example, how pixels of the screen of an Atari game change per action \cite{mnih2015human}, and thus, we have to play it to understand its dynamics.
        
        The goal of reinforcement learning is to find an \textit{optimal policy} $\pi*$ that obtains the highest possible expected return over all episodes, \textit{i.e.} $\pi* = \arg\max_{\pi} \mathbb{E} [\sum_{t=0}^T r_t]$, where the expectation goes over all initial states and all possible trajectories, which depend on the policy $\pi$. There may be more than one optimal policy. One family of solutions for this problem are so-called \textit{policy gradient methods} \cite{sutton2018reinforcement}, which is the core of our work due to their similarity with gradient-based learning used in end-to-end deep learning. Other methods, like \textit{value-based learning} (\textit{e.g.} deep Q-learning \cite{mnih2015human}), could also be used in this work but as they contain more moving components, we opt for policy gradient methods for their simplicity and previous use in supervised learning \cite{jang2017categorical, yu2017seqgan}.
        
    \subsection{Policy gradients and REINFORCE}
        Policy gradient methods use the gradient of the objective function $\mathbb{E} \left [ \sum_{t=0}^T r_t \right ]$ to perform gradient ascent to find better policies. If the policy $\pi_\theta$ is represented by parameters $\theta$ (\textit{e.g.} the weights of a deep neural network), the \textit{policy gradient theorem} \cite{sutton2000policy} states that\footnote{While \cite{sutton2000policy} derives this for state-action values (not covered here), this also works for sum of episodic returns (used here) and various other choices of value \cite{gae}.} 
        \begin{equation}
            \label{eq:pg}
            \nabla_\theta \mathbb{E} \left [ \sum_{t=0}^T r_t \right ] = \mathbb{E} \left [ \sum_{t=0}^T \nabla_\theta \log \pi(a_t | s_t) \left ( \sum_{t=0}^T r_t \right )  \right ]  
        \end{equation}
        where $\nabla_\theta \bullet$ denotes the vector of all partial derivatives of $\bullet$ w.r.t $\theta$, \textit{i.e.} the gradient. With the gradient we can then do gradient-ascent to maximize the objective $\mathbb{E} \left [ \sum_{t=0}^T r_t \right ]$ directly, \textit{i.e.} increase expected episodic reward, by updating the policy $\pi_\theta$. 
        
        Looking at the right-hand side of \eqref{eq:pg} we see that this gradient changes the probability of actions ($\log \pi$), weighted by how good the episode was (sum of rewards). If our environment only has one action per episode, the term on the right-hand side reduces to $\nabla_\theta \pi(a_0 | s_0) r_0$. Now, if the reward $r_0$ is negative, updating with this gradient \textit{will decrease the probability the taken action $a_0$ in state $s_0$}, or informally \say{\textit{do not do that}}. If the reward is positive, gradient says to \say{\textit{do it more}}. In plain words, we encourage or discourage the actions we tried, \textit{based on the reward gained}. Much like a pet dog being trained to sit on command, our policy enjoys its virtual cookies and learns to repeat actions that lead to these cookies.  

        One instance of the policy gradient methods is the REINFORCE algorithm \cite{williams1992simple}, which iteratively updates the policy parameters by estimating the gradient in \eqref{eq:pg}, and then updates the policy with gradient ascent. Using Atari game \say{Pong} as an example: the policy takes the role of the player, plays one episode of the game and either wins ($\sum_t r_t = 1$) or loses ($\sum_t r_t = -1$). We collect the states and actions policy chose during the game, and after the game we can use these in \eqref{eq:pg} to estimate the gradient and update the policy. If policy won the game, this update will encourage to take the actions it took during the single game. Likewise, if it lost, the update will discourage taking those actions. This is proven to have zero bias when estimating the gradient, but its high variance makes learning difficult \cite{sutton2018reinforcement}, which is a continuing subject of studies within reinforcement learning research (see \textit{e.g.} \cite{ppo, gae}).

\sectionpresqueeze
\section{Optimizing tandem systems with policy gradients}
\sectionpostsqueeze
\label{sec:method}
    In this section we cover the process of optimizing both ASV and CM systems working as one tandem system for a reduced t-DCF by using the REINFORCE algorithm with policy gradients.
    
    \subsection{Optimizing tandem system with REINFORCE}
        \begin{algorithm}[t]
        \label{algo:reinforce}
        \SetAlgoLined
         \KwIn{
            Pre-trained ASV $\pi_\text{asv}$ and CM $\pi_\text{cm}$ with combined parameters $\boldsymbol \theta$, dataset $D$, reward function $R$, mini-batch size $B$\;
         }
         \While{training}{
          Initialize loss $\mathcal L \leftarrow 0$ \\
          \For{$i \in \{1, 2, \ldots B\}$}{
            Sample one trial $s_\text{asv}, s_\text{cm}, \hat a_\text{asv}, \hat a_\text{cm} \sim D$ \\
            Sample actions (binary decisions). \\
            $a_\text{asv} \sim \text{Bernoulli}(\pi_\text{asv}(s_\text{asv}))$ \\
            $a_\text{cm} \sim \text{Bernoulli}(\pi_\text{cm}(s_\text{cm}))$ \\
            Combine for tandem action. \\
            $a_\text{tandem} \leftarrow a_\text{asv} \land  a_\text{cm}$ \\
            Compute probability of the tandem action. \\
            $p_\text{tandem} \leftarrow 
            \begin{cases}
                \pi_\text{asv}(s_\text{asv}) \pi_\text{cm}(s_\text{cm}),&a_\text{tandem} = 1 \\
                1 - \pi_\text{asv}(s_\text{asv}) \pi_\text{cm}(s_\text{cm}),&a_\text{tandem} = 0
            \end{cases}$ \\
            Compute reward (Section \ref{sec:reward_model}). \\
            $r \leftarrow R(a_\text{tandem}, \hat a_\text{asv}, \hat a_\text{cm})$ \\
            Accumulate policy gradient losses. \\
            $\mathcal L \leftarrow \mathcal L + \frac{1}{B} (\log p_\text{tandem}) r$ \\
           }
           Backpropagate to obtain gradient $\nabla_{\boldsymbol \theta} \mathcal L$ and update parameters $\boldsymbol \theta$ with gradient ascent.
         }
         \caption{Optimizing a tandem system with REINFORCE}
        \end{algorithm}

        The process of automatic speaker verification can be seen as a simple MDP with only two possible actions and one transition: \texttt{ACCEPT} or \texttt{REJECT}. Same applies to the countermeasure system. For each episode the system (policy) receives its input (state) and decides on whether to accept or reject the trial (take an action). It is then rewarded depending on whether the decision was right or not. In the case of multiple systems working in tandem (ASV and CM), we can model them as one big policy that provides one final action. The final action is a logical-and of the individual actions, corresponding to the \textit{parallel} mode of system combination described in \cite{kinnunen2018t}. With this setup we can now update both systems with gradient ascent by applying the REINFORCE algorithm, where policy gradient theorem \eqref{eq:pg} allows us to backpropagate through the accept/reject decisions and logical-and operation.
        
        The training procedure is detailed in the Algorithm \ref{algo:reinforce}. The ASV and CM systems $\pi_{\text{asv}}$ and $\pi_{\text{cm}}$ map their respective inputs $s_{\text{asv}}$ and $s_{\text{cm}}$ to probabilities for accepting the trial, denoted here by $p(\text{ASV accepts trial}) = \pi_{\text{asv}}(s_{\text{asv}})$. The individual actions $a_\text{asv}, a_\text{cm} \in \{0, 1\}$ are then sampled individually from the \textit{Bernoulli distributions} with the respective probabilities, and the tandem action $a_\text{tandem} \in \{0, 1\}$ for accepting the trial is a logical-and of the individual actions. In the process we obtain the probability of the tandem action $p_\text{tandem}$. With the ground-truth labels $\hat a_\text{asv}$ and $\hat a_\text{cm}$ we can then assign a reward $R(a_\text{tandem}, \hat a_\text{asv}, \hat a_\text{cm})$ to this episode, discussed below. These rewards are then used to accumulate losses that reflect policy gradient \eqref{eq:pg} over a mini-batch. Finally, we use the automatic differentiation tools of PyTorch \cite{pytorch} to compute gradients of the accumulated loss and apply gradient descent to update \textit{all} parameters of both systems.
        
        The major thing to note here is the \textit{sampling of actions}: instead of a fixed threshold like done in typical ASV systems, the systems' actions are randomly picked according to probabilities they output\footnote{This is a referred to as \textit{stochastic policy}. A thresholded ASV system would be a \textit{deterministic policy}, a special case of policy which is used in some of the reinforcement learning methods (\textit{e.g.} Deep Q-learning \cite{mnih2015human}).}. \textit{However}, in evaluation and final use-case we would still use a fixed threshold, and sampling is \textit{only} used during the training phase. The motivation for this is clearer from reinforcement learning side: if we always play the same move in same situation while learning chess, we never \textit{explore} other strategies that would be available after taking different moves. On the other hand, if we always try different actions, we never \textit{exploit} the knowledge we already have and explore already known strategies further. By sampling actions, as in Algorithm \ref{algo:reinforce}, we balance between exploration and exploitation. This too is an active field of research in reinforcement learning \cite{sutton2018reinforcement, pathak2017curiosity}. 

    \subsection{Reward functions}
    \label{sec:reward_model}
        We still have to decide our reward function $R(a_\text{tandem}, \hat a_\text{asv}, \hat a_\text{cm})$, which defines the optimization target. We include three simple reward functions commonly used in the reinforcement learning, and one derived from the t-DCF cost-model.

        \textbf{Simple}: Arguably the simplest solution is $R(\bullet) = 1$ reward for success (correct decision) and $R(\bullet) = -1$ reward for failure (incorrect decision). This is a common reward model in reinforcement learning, and it was used to define the goal of winning the opponent while mastering Go \cite{silver2016mastering} and Starcraft II \cite{vinyals2019grandmaster} with reinforcement learning.
        
        \textbf{Reward}: Another common reward model is to reward only success with $R(\bullet) = 1$ and give no explicit penalty for failure. This reward model is common in tasks where there is no explicit way to fail, \textit{e.g.} navigating a maze to find a goal \cite{pathak2017curiosity}. For our purposes this may not be ideal: incorrect actions do not lead into any updates (zeroed out by the reward), and positive reward of correct action keeps pushing those scores higher/lower without limits, which may lead to overfitting in terms of very sharp predictions. If the initial systems often picks incorrect actions, this would also lead to small number of samples with non-zero gradients and thus noisy updates and/or slow learning.
        
        \textbf{Penalize}: More intuitive solution for our purposes is only penalizing incorrect decisions with $R(\bullet) = -1$, and with no reward for correct decisions. Think of a deterministic case with distribution of target scores, nontarget scores and a fixed threshold. If a sample is on the wrong side of the threshold, this reward model would push it towards the correct side. If sample is already on the right side, nothing happens. Meanwhile \textit{reward} model would do the opposite: move correct scores further away from threshold, but not move the incorrect samples towards the other side of the threshold. For this reason, we expect this reward function to work better than \textit{reward}.

        \textbf{t-DCF}: Previous reward functions separate only between the correct and incorrect decisions, but we can also define rewards based on individual ground truths, as done in the t-DCF described in Section \ref{sec:tdcf}. This is where we optimize more directly for t-DCF by using its cost parameters. Recall that REINFORCE operates on \textit{individual} trials, and for each sample we compute separate reward. If we compute the t-DCF \eqref{eq:new_tdcf} of an individual trial, at most one of the $P_\bullet$ term remains with value of one, along with the corresponding $C_\bullet$ and $\rho_\bullet$ terms. Repeat this for all the possible combinations of trials (target, nontarget and spoof) and possible predictions, and we obtain t-DCFs for individual trials. These rewards (negative of t-DCF) are summarized in the Table \ref{table:tdcf_reward}. This reward function is similar to \textit{penalize} reward function with parameters that allow weighting different error situations differently.

        \begin{table}[]
            \footnotesize
            \centering
            \begin{tabular}{cc|ccc}
            \textbf{ASV}                     & \textbf{CM}     & \textbf{Target}                                 & \textbf{Nontarget}                           & \textbf{Spoof}                                        \\ \hline
            \multirow{2}{*}{Accept} & Accept & 0                                      & $-C_\text{fa} \rho_{\text{non}}$ & $-C_\text{fa,spoof} \rho_{\text{spoof}}$ \\
                                    & Reject & $-C_\text{miss} \rho_{\text{tar}}$ & 0                                    & 0                                            \\
            \multirow{2}{*}{Reject} & Accept & $-C_\text{miss} \rho_{\text{tar}}$ & 0                                    & 0                                            \\
                                    & Reject & $-C_\text{miss} \rho_{\text{tar}}$ & 0                                    & 0                                           
            \end{tabular}
            \tablesqueezevspace
            \caption{{\it The reward-function derived from t-DCF cost function. Note that there are no positive rewards, only penalties.}}
            \label{table:tdcf_reward}

        \end{table}
        
        Note that all of the above reward functions have the same goal and thus the same optimal policy: \say{predict the correct accept/reject decision no matter the input}. In that case, does it make sense to define different reward functions? In reinforcement learning this type of \textit{reward shaping} \cite{ng1999policy} can be used to speed up the learning process, depending on the environment. In our scenario the target/nontarget samples could be inseparable, due to choice of system architectures, features or data alone. If so, the reward function would bias training towards rejecting spoof samples than overly accepting, if the cost of false-accept is higher than false-reject, for example.
    
    \subsection{Related work}
        REINFORCE, or also called \say{score function estimator} \cite{jang2017categorical}, is no stranger to supervised learning. It has been used to \textit{e.g.} compute gradient through loss metric that requires generating a sequence of characters \cite{yu2017seqgan}. The version of REINFORCE used in this work is the simplest version, but multiple alternatives have been proposed over the years with stabler learning, summarized by \cite{jang2017categorical} along with their proposed method (differentiable softmax operation). 
        
        The reinforcement learning model used here for the tandem ASV-CM setup and training is not the only one available, either. One could, for example, model the situation as a \textit{multi-agent} setup with multiple individual policies \cite{boutilier1996planning}, or use the \textit{cascaded} setup where CM first has to accept the trial before passing it to the ASV \cite{kinnunen2018t}. Our focus is on studying the applicability of REINFORCE on optimizing such tandem systems in terms of t-DCF, and thus opt for the simplest approach.

\sectionpresqueeze
\section{Experimental setup}
\sectionpostsqueeze
    Here we cover the corpora, features (front-end), neural network architectures and training procedures used in the experiments. Before tandem-training we pre-train the ASV and CM systems separately. We repeat all experiments three times and report averages. This is common in reinforcement learning as stochastic nature of \textit{e.g.} REINFORCE training often leads to different results between runs \cite{henderson2018deep}. Experiment code is available at \url{https://github.com/Miffyli/asv-cm-reinforce}.

    \subsection{Corpora and features}
        
        \begin{table}[t]
            \centering
            \begin{tabular}{l|cccc}
            \textbf{Partition}& \textbf{Male} & \textbf{Female} & \textbf{Bona fide} & \textbf{Spoof} \\ \hline
            Train    & 8             & 12      & 2,580      & 22,800 \\
            Dev      & 8             & 12      & 2,548      & 22,296 \\
            Eval     & 30            & 37      & 7,355      & 63,882 \\
            \end{tabular}
            \tablesqueezevspace
            \caption{{\it Statistics of the ASVSpoof19 corpus (logical access) for training spoofing countermeasure systems, with the number of speakers by gender.}}
            \label{table:asvspoof_stats}
        \end{table}
        
        Main corpus used in this work is ASVSpoof19 dataset (logical access scenario) \cite{todisco2019asvspoof}, which provides labels for speaker verification and spoof samples, generated using different techniques. We will use ASVSpoof19 corpus for training countermeasure systems and for the tandem training later. Statistics of the portion used for training independent ASV and CM is summarized in the Table \ref{table:asvspoof_stats}.
        
        ASVSpoof19 does not contain enough data to train a reliable speaker verification system based on deep neural networks, which is why we include VoxCeleb1 \cite{Nagrani17voxceleb} data for pre-training it. We use the training and trial list provided by the authors to train our system. This list contains 1,211 individual speakers with 148,642 utterances for training, and additional 40 speakers and 4,874 utterances for testing. For each utterance we extract a x-vector with Kaldi using a pre-trained model\footnote{Model downloaded from \url{http://kaldi-asr.org/models/m7}} \cite{snyder2018x}. All the following speaker verification systems use these x-vectors of size $512$ as an input.
        
        Following ASVSpoof2019 challenges' baseline results with CQCC features \cite{todisco2017constant}, we extract CQCC features for all the utterances in the ASVSpoof2019 dataset. We use the same set of parameters as used by the ASVSpoof19 CQCC baseline system\footnote{ \url{https://www.asvspoof.org/asvspoof2019/asvspoof2019_evaluation_plan.pdf}}. The features for the CM system are thus feature matrices of $N \times 60$, $N$ depending on the utterance length. Note that, for better performance in this corpus, ASVSpoof19 challenge participants experimented with multiple different features with improved results \cite{lavrentyeva2019stc, chettri2019ensemble}. CQCC features alone may fall short in terms of raw performance, but they are readily available and designed for spoof detection \cite{todisco2017constant}.
    
    \subsection{Evaluation metrics}
        We evaluate tandem system using the normalized minimal t-DCF with the same cost parameters and form as in the ASVSpoof19 challenge \cite{todisco2019asvspoof}. We fix the ASV threshold to its \textit{equal error rate} (EER), and sweep over CM thresholds for minimal, normalized t-DCF. This t-DCF value ranges from zero (perfect system) to one (CM accepts all spoof samples). Note that this t-DCF is \textit{only} used for evaluation, not for tandem optimization. Performance of individual systems are evaluated with EER on their respective tasks. Unless otherwise mentioned, all t-DCF and EER values are computed over ASV protocols of the ASVSpoof19 dataset (trials in \texttt{ASVspoof2019.LA.asv.dev.gi.trl.txt}\footnote{Available in the ASVSpoof19 corpus at \url{https://datashare.is.ed.ac.uk/handle/10283/3336}}), including the performance of the CM system.

    \begin{figure}[t]
        \includegraphics[width=\columnwidth]{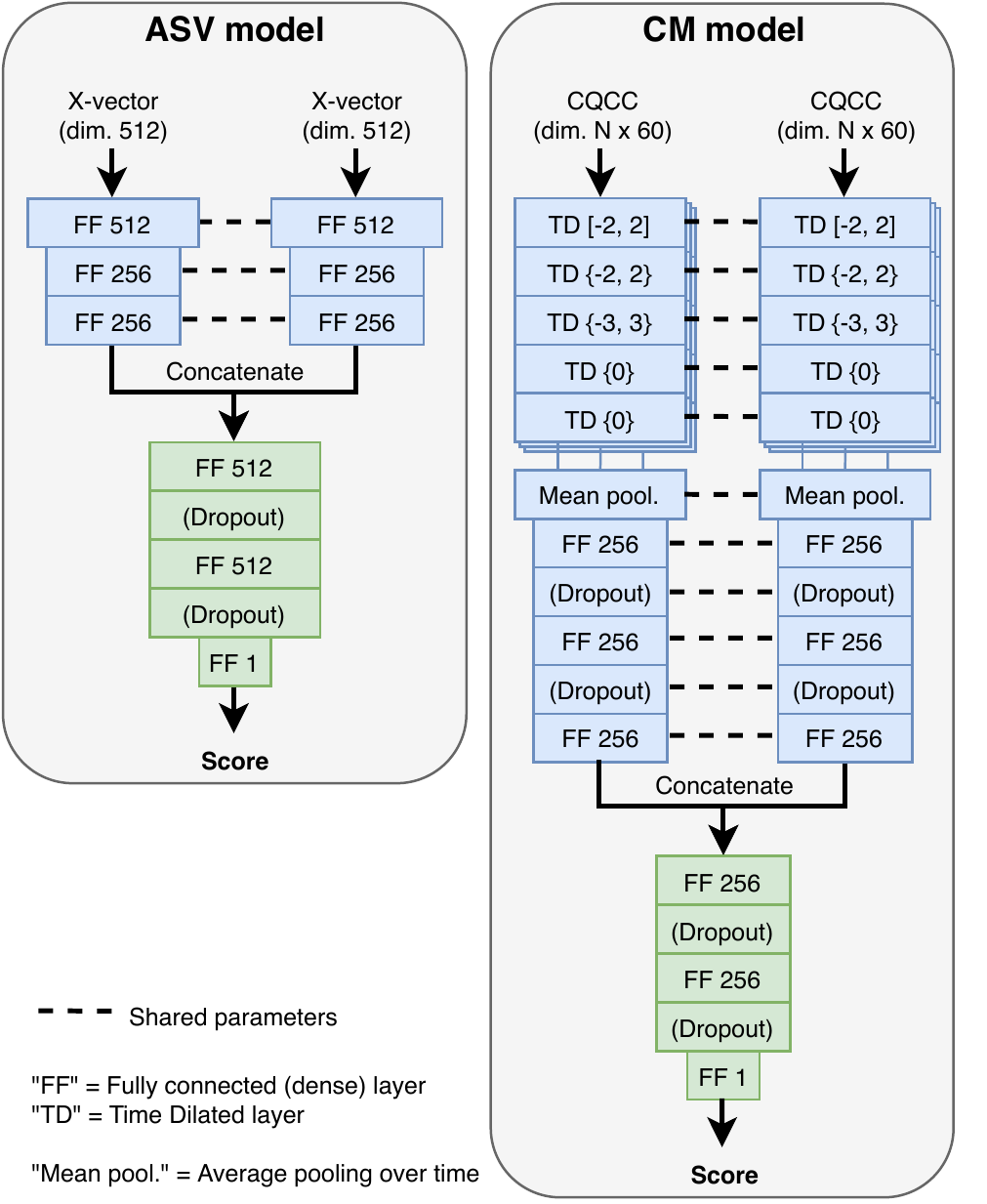}
        \figsqueezevspace
        \caption{{\it Neural network architectures for ASV (left) and CM (right) models. All hidden layers use ReLU activations. Blue boxes represent ``siamese" modules and green ``discriminator" modules. ASV system is a classification back-end with fixed embeddings, while CM system contains both front and back-ends. Time dilated layers follow the notation of \cite{peddinti2015time}, and CM model is a lighter version of X-vector model described in \cite{snyder2018x}.}}
        \label{fig:models}
    \end{figure}

    \subsection{Speaker verification system}
        The deep neural network for ASV is presented in the Fig.~\ref{fig:models}, which follows the architecture presented in \cite{ravanelli2019learning}, where authors used mutual-information maximization to train the network. All training procedures only update this back-end of speaker verification (siamese and discriminator module), while the front-end x-vectors remain fixed. Each training sample consists of two input features and a $\{0, 1\}$ target. If both features (x-vectors) originate from the same speaker, the target is $1$, otherwise $0$. Vectors are sampled from the VoxCeleb1 speaker verification list \cite{Nagrani17voxceleb}, and the network is updated with mini-batches of size $64$, each containing equal amount of samples with targets $0$ and $1$ to avoid class-imbalance in the training. The network parameters are updated to minimize the cross-entropy loss between sigmoided output of the network and the target labels  using Adam \cite{adam}, with learning rate of $0.001$ for $60$ epochs, and with L2-regularization weight of $5 \cdot 10^{-5}$ and no dropout. These hyper-parameters were chosen through a grid-search on the development set of ASVSpoof19 over L2-regularization weights $\{0, 10^{-5}, 5 \cdot 10^{-5}, 10^{-4}\}$, dropout rates $\{0, 0.25, 0.5, 0.75\}$ and different training lengths of $[5, 60]$ epochs. We observed no benefit from increasing layer widths.

        After the training on VoxCeleb1 corpus, we adapt to ASVSpoof19 corpus by fine-tuning parameters with the CM training list (\texttt{ASVspoof2019.LA.cm.trn.txt}) for $20$ epochs, using Adam with learning rate $0.0001$. This reduced the relative EER by $24\%$ on the ASVSpoof19 evaluation set. The three repetitions reached EERs of $9.74\%$, $10.73\%$ and $8.90\%$ in the evaluation set. This is a high error-rate compared to \textit{e.g.} system used in the ASVSpoof19 challenge which had an EER of $2.28\%$ with x-vectors and PLDA back-end \cite{todisco2019asvspoof}. However, our focus is not to reach low error-rates, rather to have an initial system which we then further optimize later with reinforcement learning.

    \subsection{Spoofing countermeasure system}
        CM system uses similar network as ASV, presented in the Fig.~\ref{fig:models}. The siamese module uses an architecture similar to the x-vector network \cite{snyder2018x}, but with smaller layers. All training procedures update both the front-end (siamese) and back-end (discriminator). We use the same mutual-information maximization training \cite{ravanelli2019learning} as for ASV: if both two sampled utterances are bona fide samples, the target is $1$, otherwise it is $0$. No L2-regularization or dropout was used, and training was halted after $10$ epochs. These hyper-parameters were also chosen by the same grid-search on ASVSpoof19 development set, with range of parameters in the case of ASV system. For scoring evaluation trials, we average the embedding features of all bona fide samples and use it as the second input when testing if trial sample is bona fide sample or not. Embedding vector is obtained from the output of the siamese module. We observed a slightly lower average performance without the siamese architecture, in both CM and tandem training.
        
        The three repetitions reached EERs of $7.25\%$, $11.75\%$ and $7.06\%$ on the evaluation set. The high error-rate of the second repetition indicates that the training of this system is sensitive to when exactly training is stopped. We speculate this is due to the \say{average bona-fide vector} described above, instead of using a more common network with one input and a single scalar output for classification.
    
    \subsection{Tandem training}
        \begin{figure}[t]
        \centering
            \includegraphics[width=1\columnwidth]{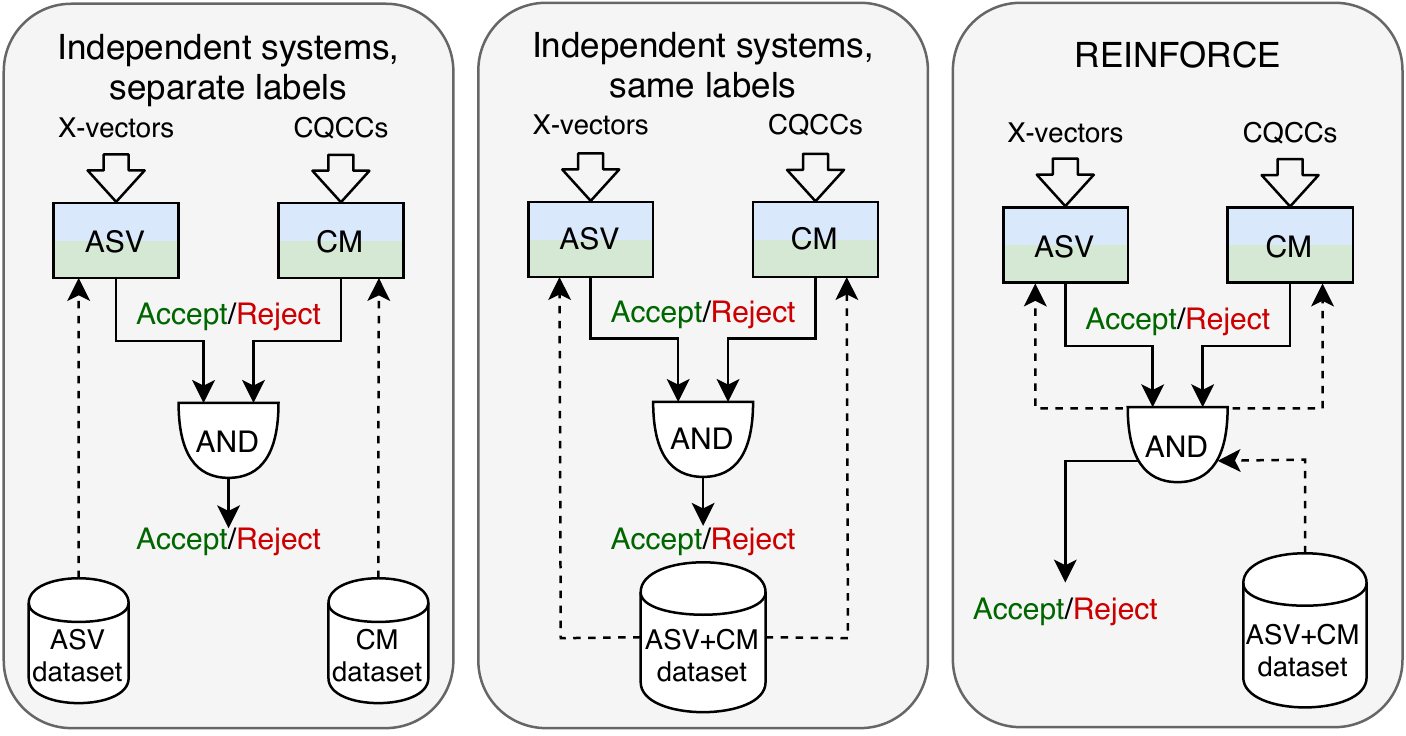}
            \figsqueezevspace
            \caption{{\it Methods for optimizing tandem systems explored in this work, along with the baseline where the systems are only trained separately. ``ASV+CM dataset" refers to dataset with only tandem accept/reject labels. Dashed lines represent flow of training information (i.e. backpropagation).}}
            \label{fig:methods}
        \end{figure}

        With pre-trained systems we combine them in parallel fashion to create a spoof-robust speaker verification, which we then optimize. This optimization affects all the parameters represented in Fig.~\ref{fig:models} (back-end for ASV, front and back-end for CM).  Our main focus is on REINFORCE-based tandem optimization detailed in Section \ref{sec:method}, which we will study more in depth experimentally and compare against baseline methods. All methods explored here are summarized in Fig.~\ref{fig:methods}.
        
        All methods use ASVSpoof19's development set's gender-independent trial list for the tandem-optimization (\texttt{ASVspoof2019.LA.asv.dev.gi.trl.txt}), which contains the necessary ASV and CM labels for each trial. This file list contains 10 different speakers with 2,548 bona fide and 22,296 spoof samples\footnote{Note that this is a separate file list from countermeasure partitions listed in the Table \ref{table:asvspoof_stats}.}. Training batches are constructed as in the pre-training phase, described in the previous sections. To avoid catastrophic forgetting and/or destructive updates due to high variance of the REINFORCE training \cite{jang2017categorical}, all the methods use stochastic gradient descent with learning rate $0.0001$ and mini-batch size $64$, with total of $50$ epochs of training. Each mini-batch contains roughly equal amount of target and nontarget sample.

        \textbf{Independent models, separate labels.}
            The pre-trained ASV and CM systems were not trained with the data available in this tandem-training phase. For a fair comparison with REINFORCE-based method we fine-tune the pre-trained ASV and CM with their respective labels using cross-entropy loss. Note that there is no tandem training here, we merely use the new data available to tune the systems. This will be abbreviated \say{IM-separate}.
        
        \textbf{Independent models, same labels.}
            Similar to the fine-tuning above, we train the ASV and CM systems with cross-entropy loss, but this time the target labels represent tandem decision (target only if sample is bona fide \textit{and} speaker is target speaker). That is, both systems are trained to do the task of the other one, with ASV learning to discard spoof samples and CM learning to reject nontarget speakers. We believe that with a slow learning rate this keeps systems in the range of their original tasks, but also utilizes their unique models/features for the other task. ASV's x-vector embeddings could contain some features that make it easy to distinguish synthetic speech, for example. This is will be abbreviated \say{IM-same}.
        
        \textbf{REINFORCE.}
            For the REINFORCE method described in Section \ref{sec:method}, we include experiments with the four explained reward functions \textit{simple}, \textit{penalize}, \textit{reward} and \textit{t-DCF}. For \textit{t-DCF} reward we use the same cost-parameters used in the evaluation: $C_\text{miss} = 1$, $C_\text{fa} = C_\text{fa,spoof} = 10$, $\rho_\text{tar} = 0.95 \cdot 0.99, \rho_\text{non} = 0.95 \cdot 0.01$ and $\rho_\text{spoof} = 0.05$.

\begin{figure*}[t]
    \newcommand{\hspacing}{\hspace{-0.5cm}}
    \newcommand{\figwidth}{0.185\textwidth}
    \centering
    \subfigure{\includegraphics[width=\figwidth]{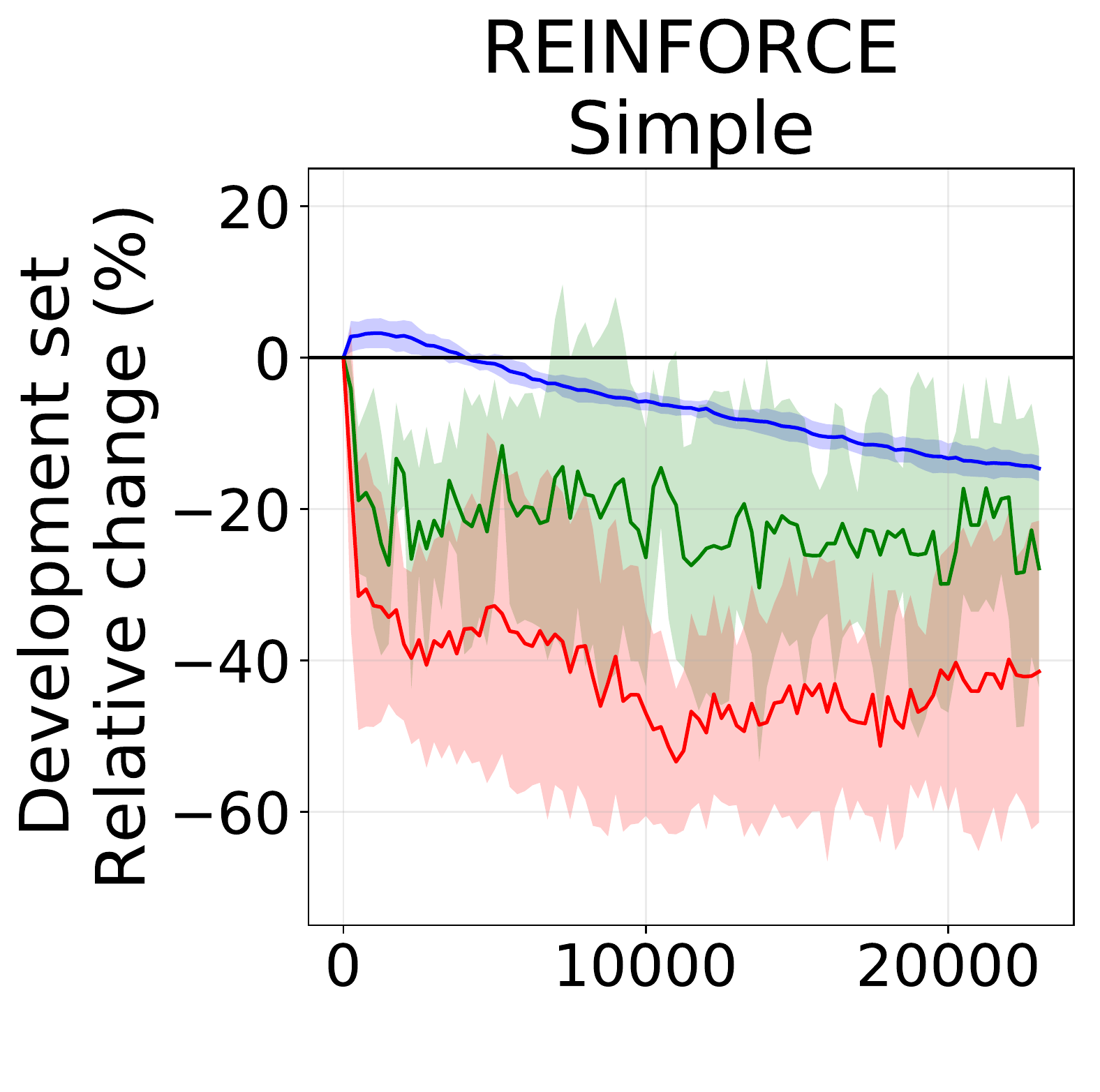}}
    \hspacing
    \subfigure{\includegraphics[width=\figwidth]{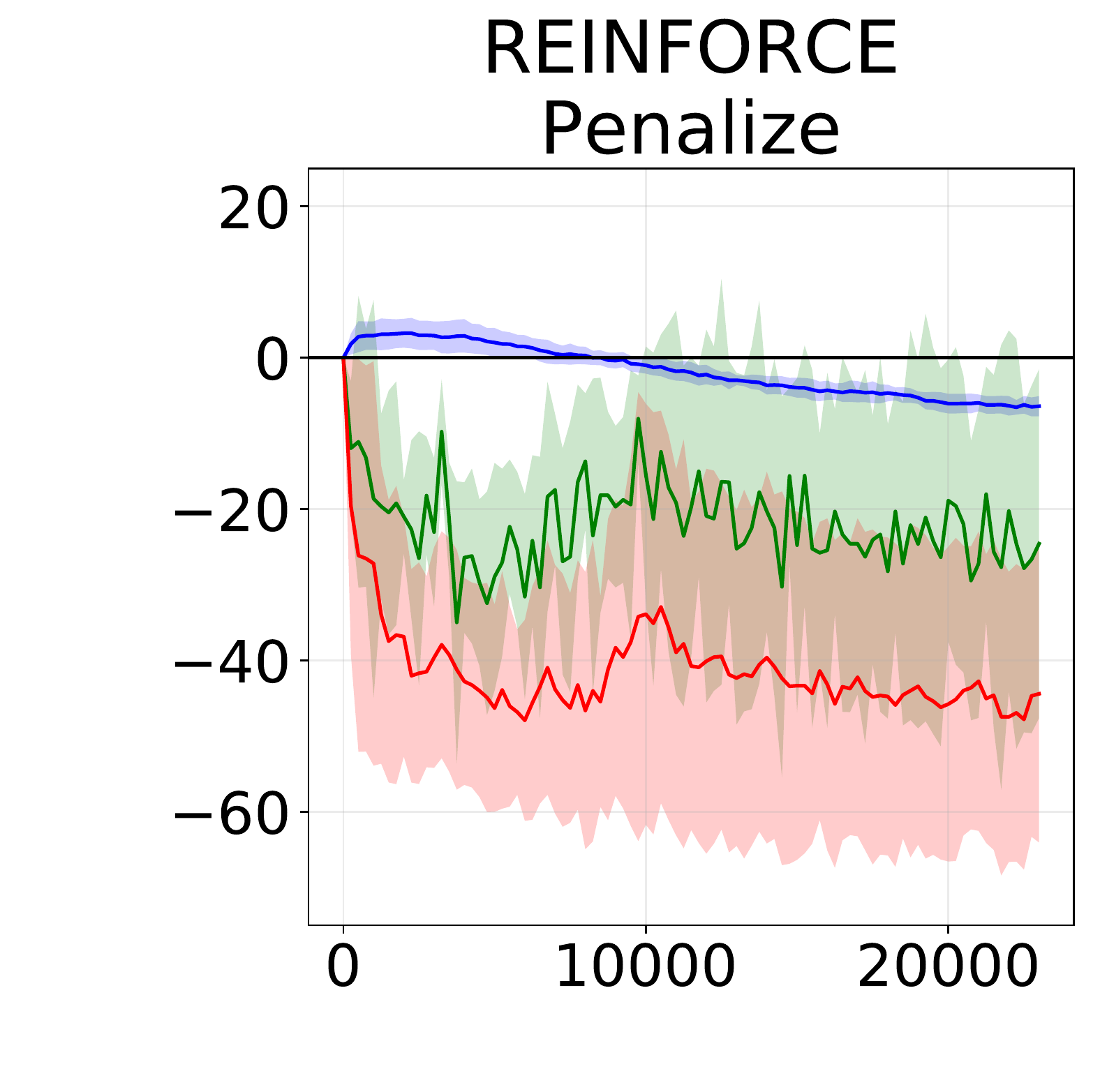}}
    \hspacing
    \subfigure{\includegraphics[width=\figwidth]{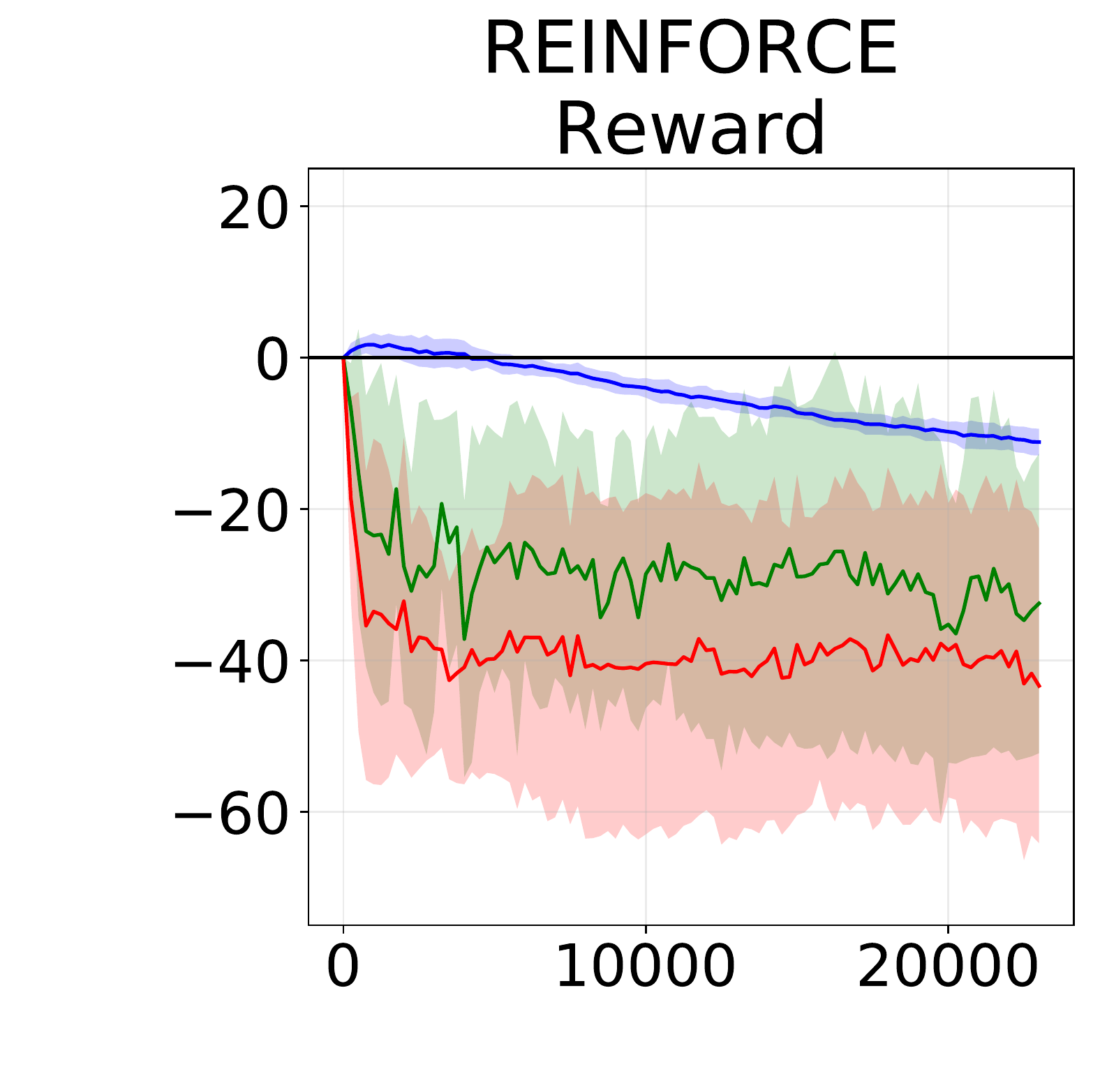}}
    \hspacing
    \subfigure{\includegraphics[width=\figwidth]{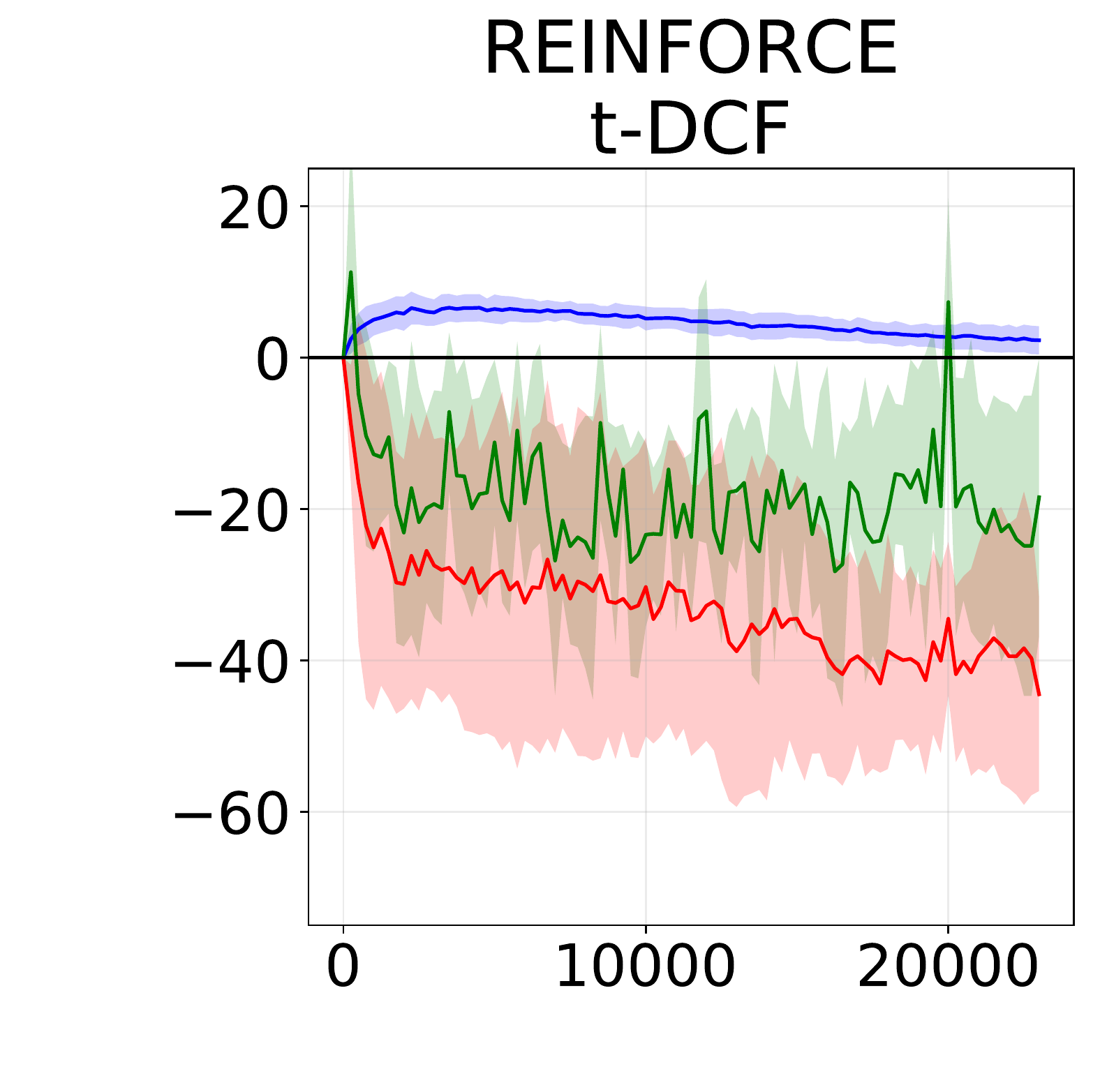}}
    \hspacing
    \subfigure{\includegraphics[width=\figwidth]{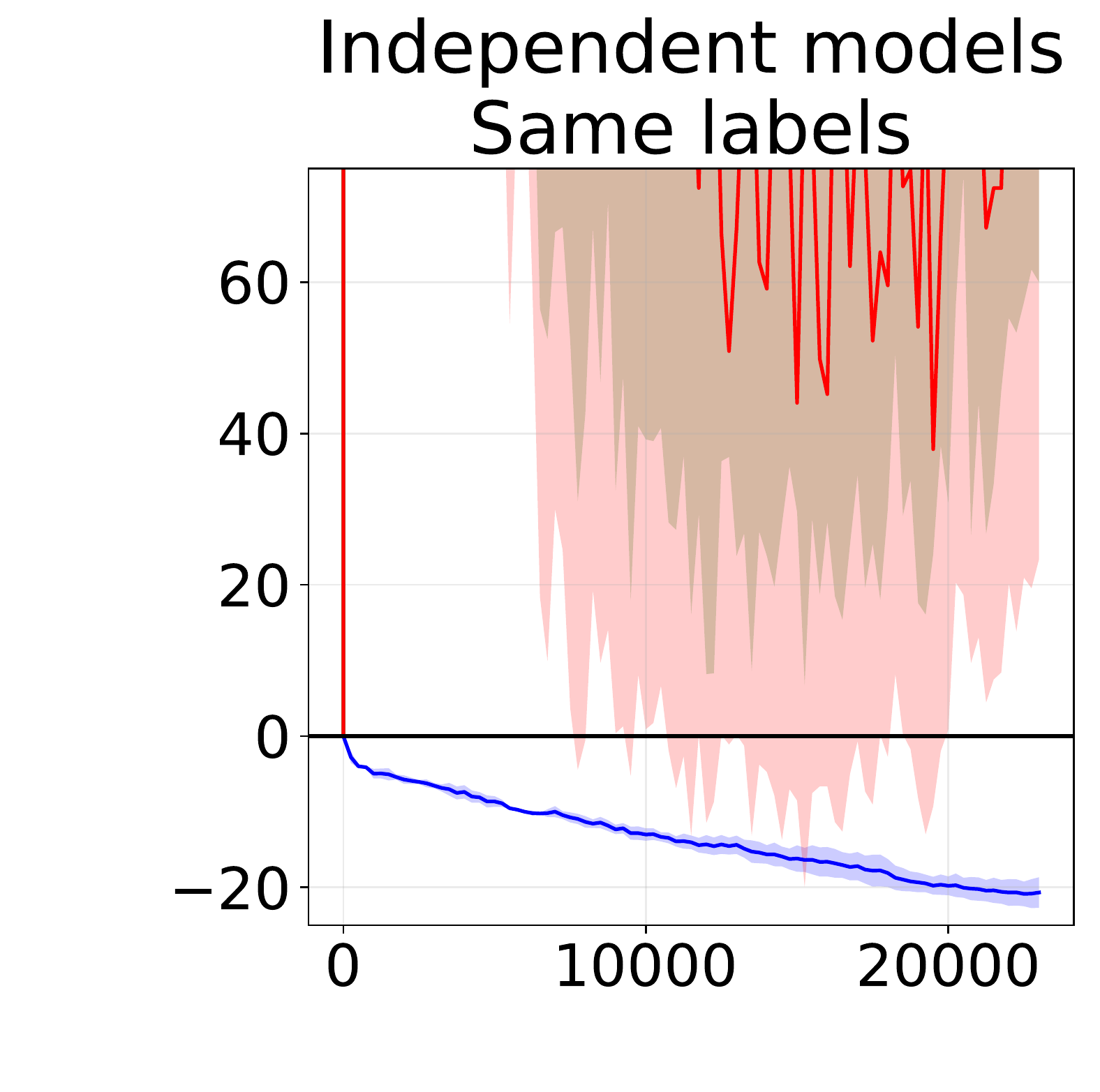}}
    \hspacing
    \subfigure{\includegraphics[width=\figwidth]{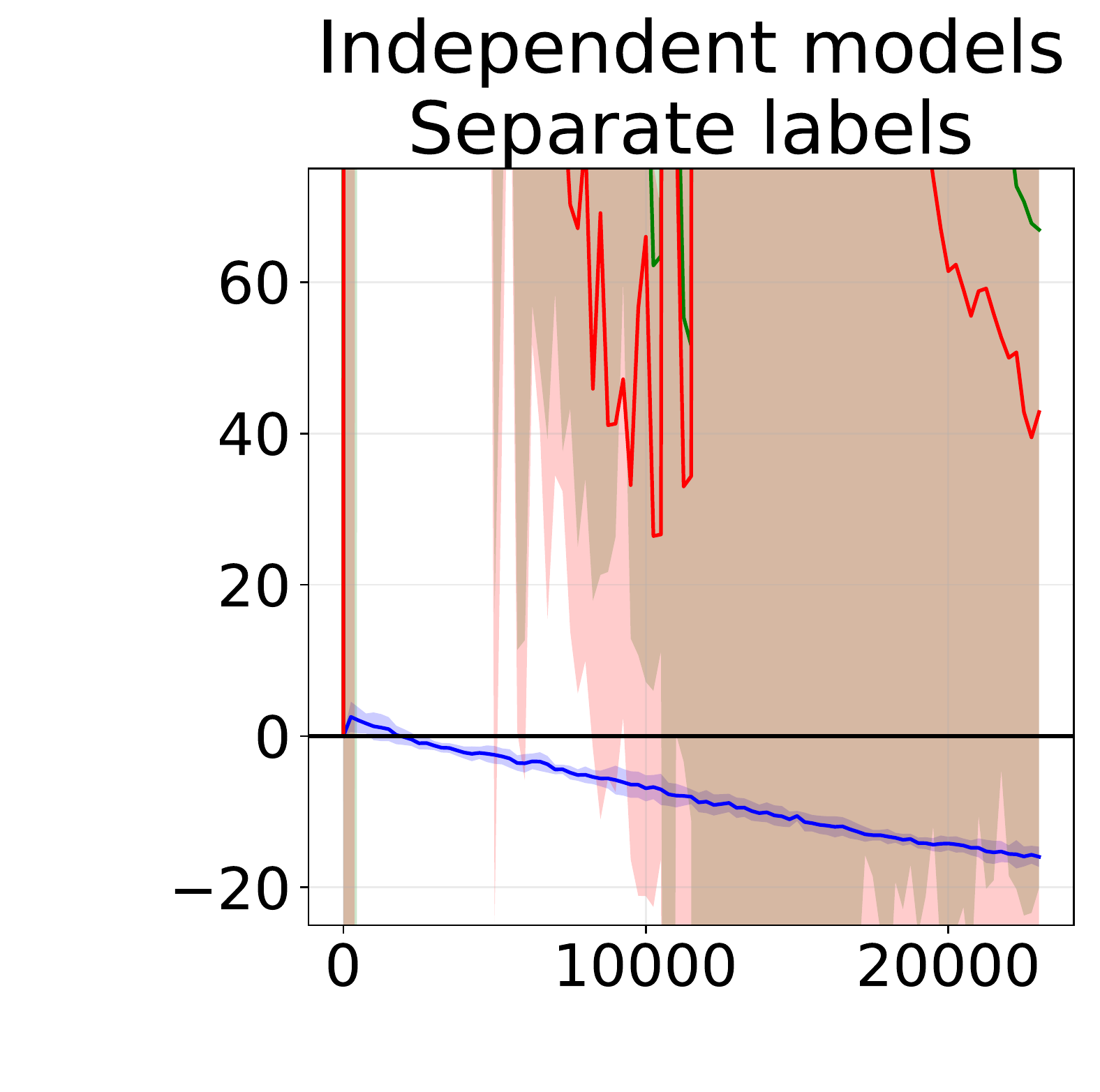}}
    \\ \vspace{-.6cm} 
     \hspace{0.0cm}
    \subfigure{\includegraphics[width=\figwidth]{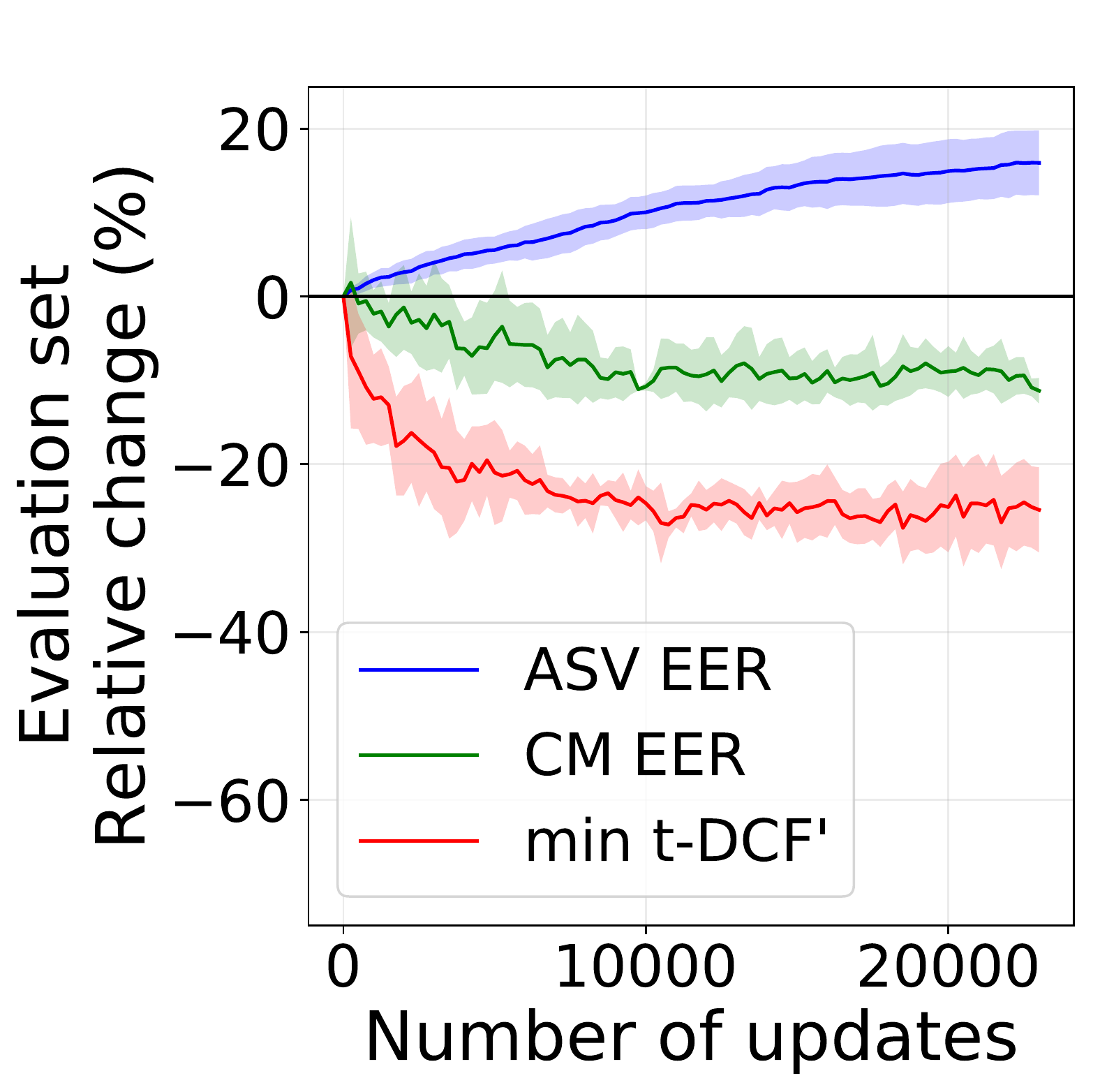}}
    \hspacing
    \subfigure{\includegraphics[width=\figwidth]{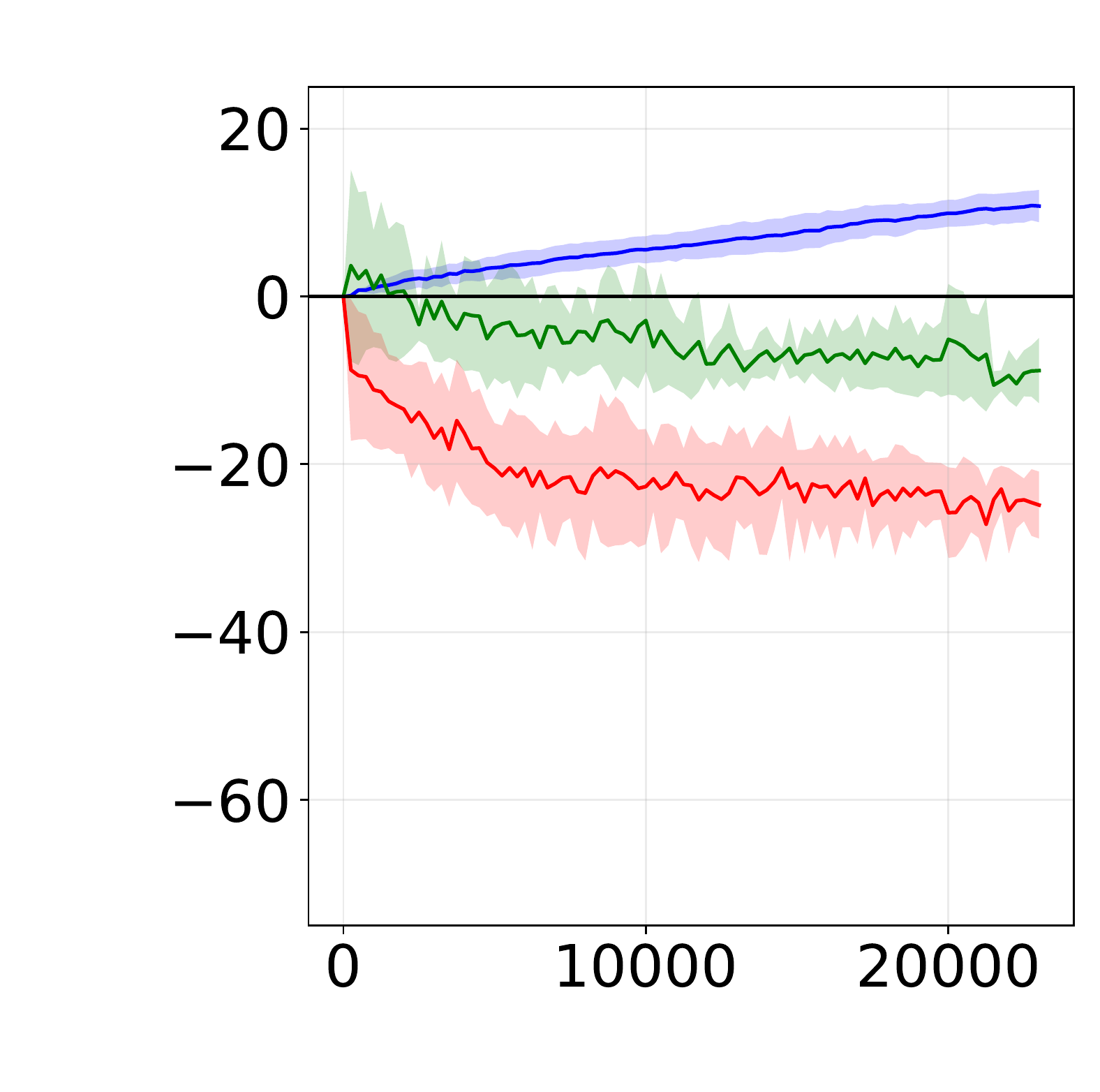}}
    \hspacing
    \subfigure{\includegraphics[width=\figwidth]{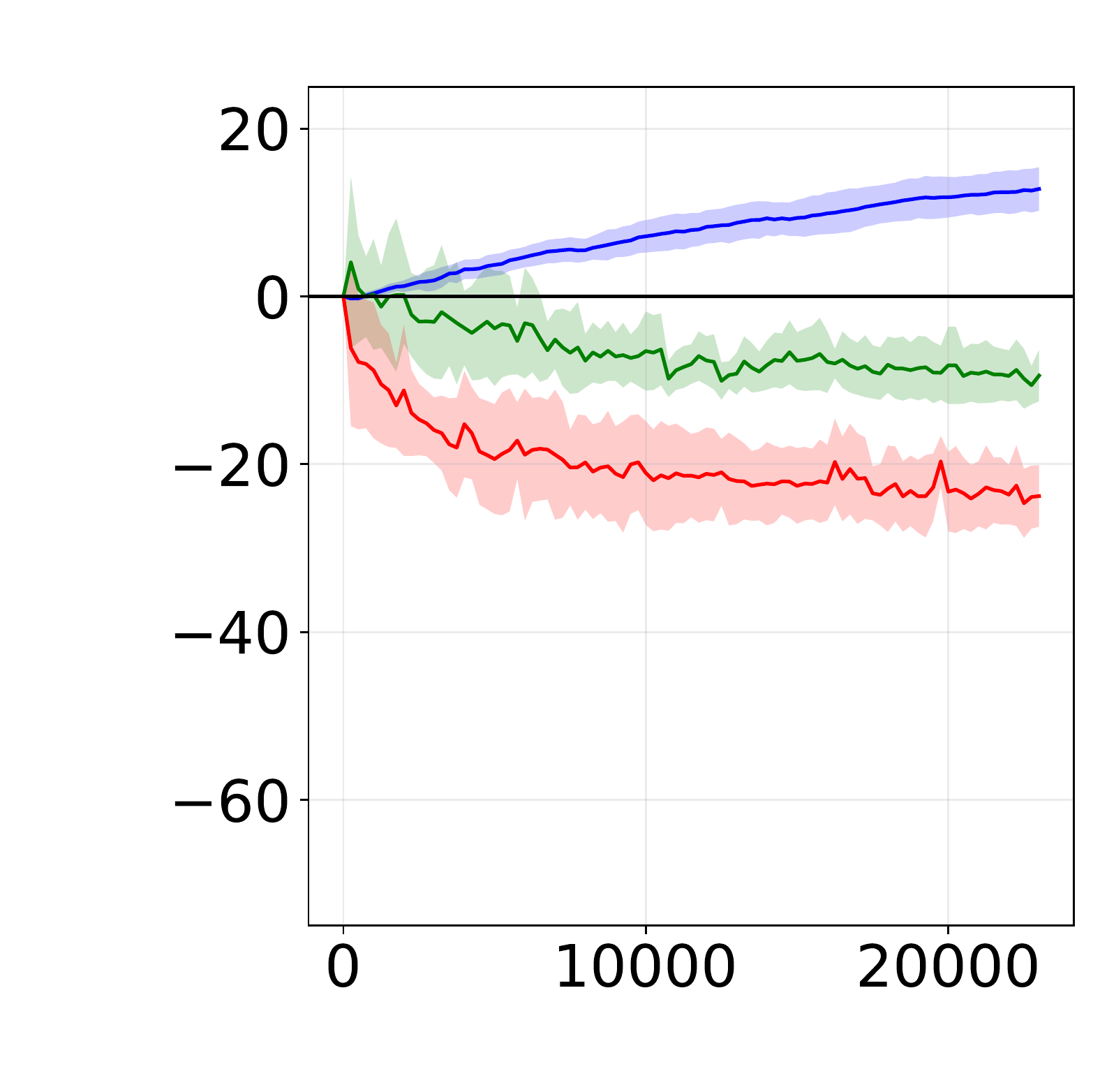}}
    \hspacing
    \subfigure{\includegraphics[width=\figwidth]{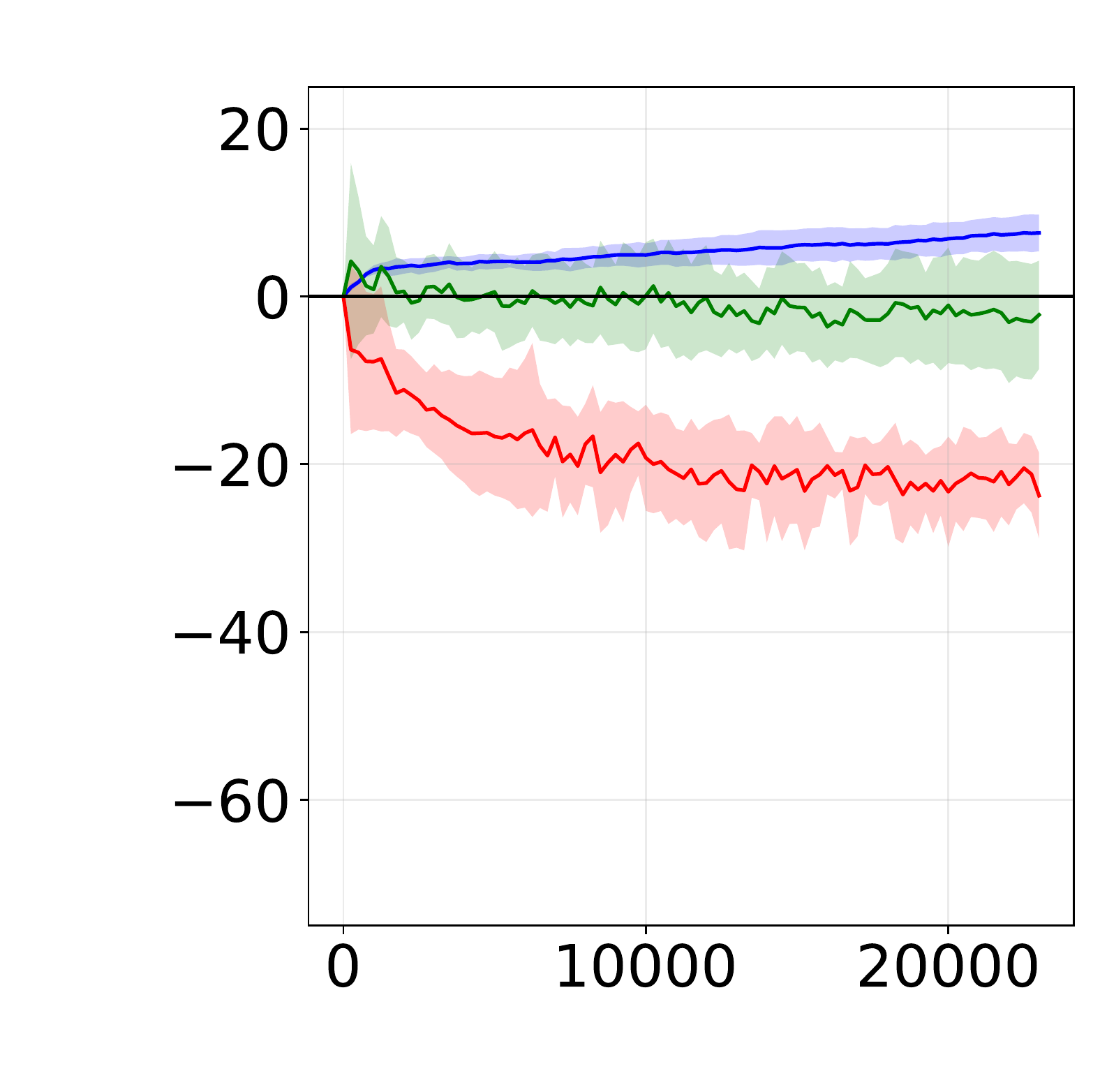}}
    \hspacing
    \subfigure{\includegraphics[width=\figwidth]{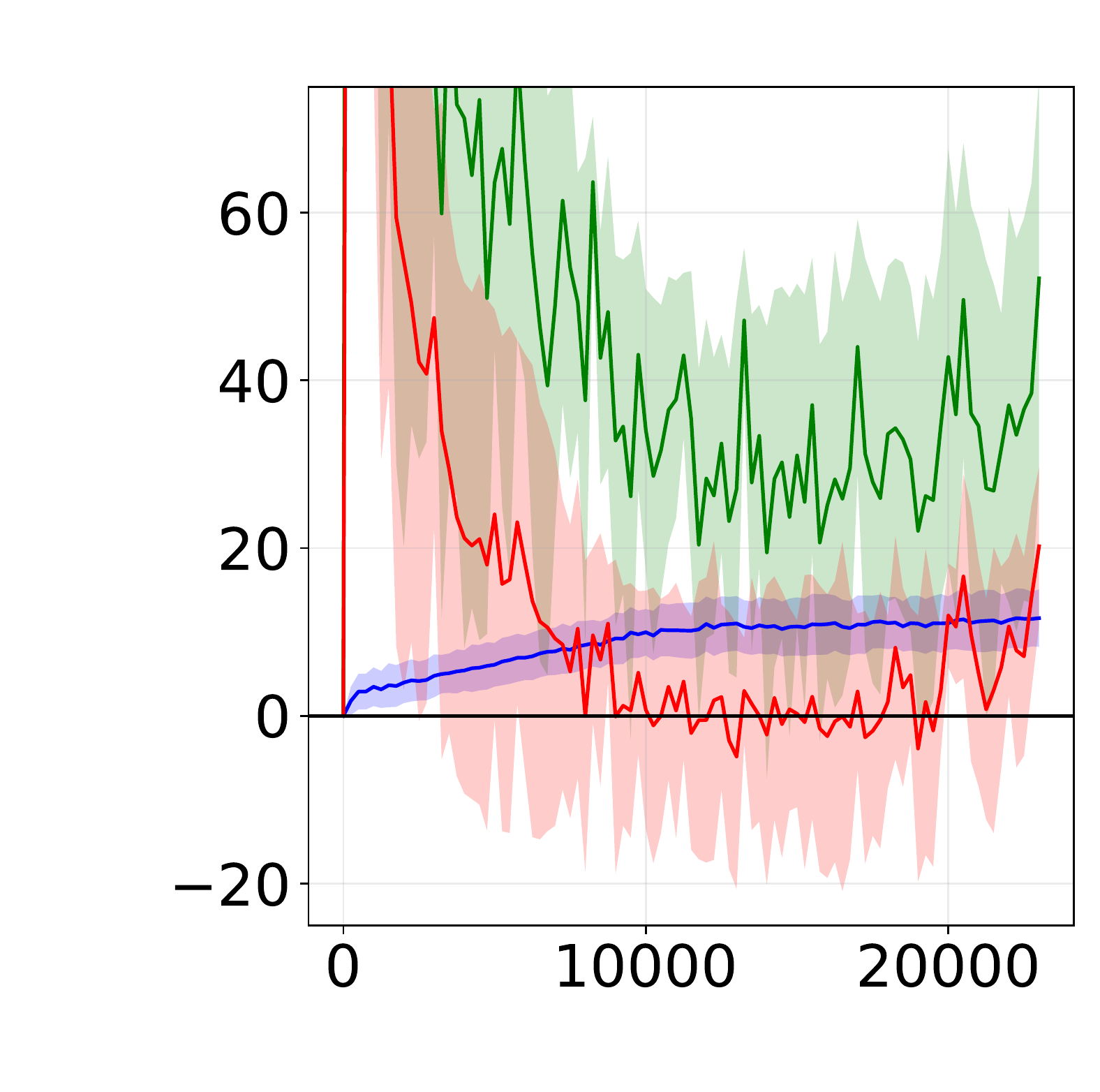}}
    \hspacing
    \subfigure{\includegraphics[width=\figwidth]{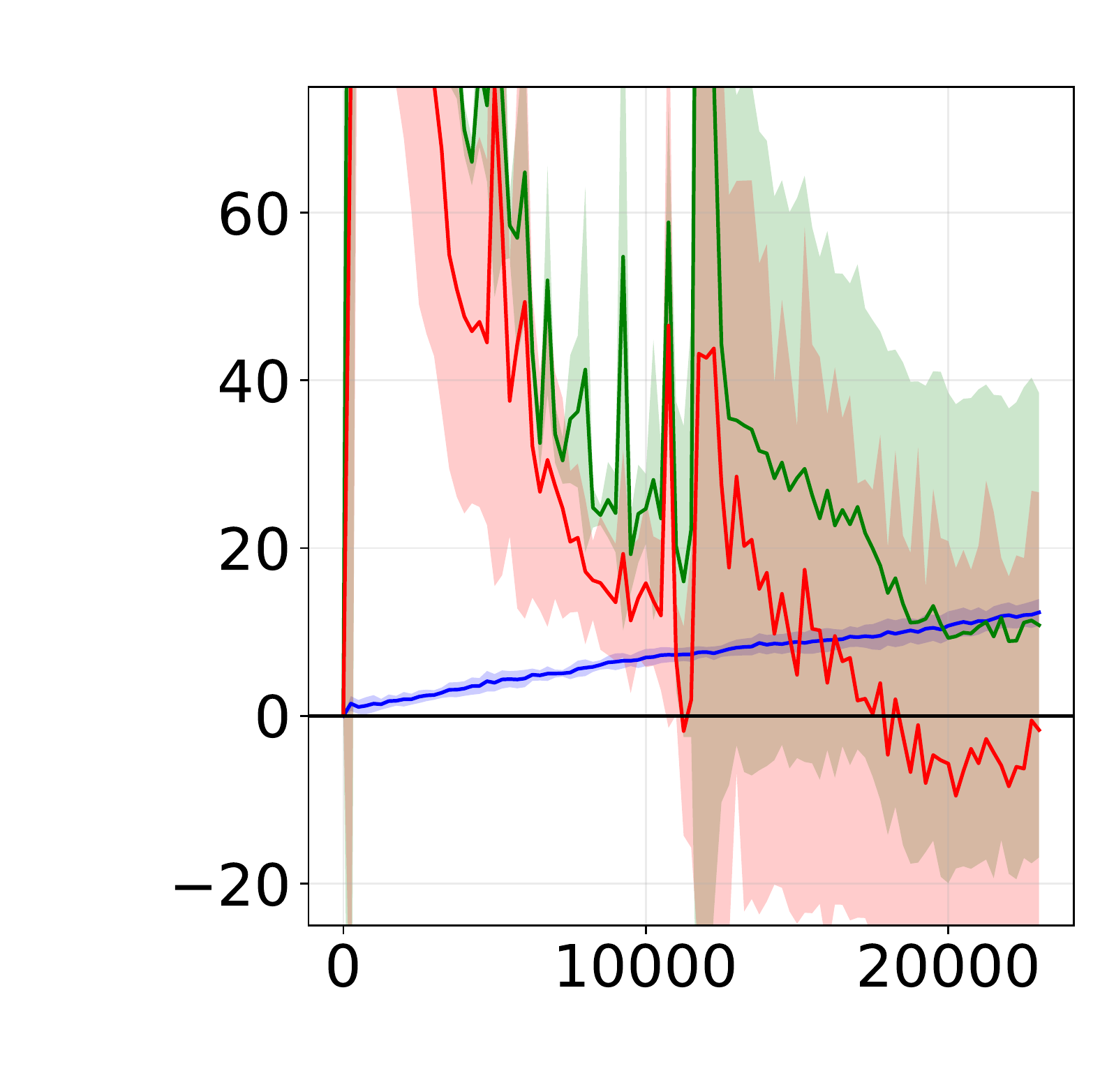}}
    \figsqueezevspace
    \caption{{\it Performance metrics over tandem optimization with REINFORCE, evaluated on development (training) and evaluation sets (first and second row, respectively). Y-axis shows the change in the metric relative to the model before tandem optimization. Curves are averaged over three experiments, with shaded region representing standard deviation. REINFORCE training reliably improves t-DCF over training, while cross-entropy (``Independent models") based methods cause spikes in the error-rates}}
    \label{fig:learning_curves}
    
\end{figure*}
    
\sectionpresqueeze
\section{Results and Discussion}
\sectionpostsqueeze    
    \begin{table}[t]
        \centering
        \footnotesize
        \begin{tabular}{c|cccc}
        & \multicolumn{2}{c}{\textbf{Development}} & \multicolumn{2}{c}{\textbf{Evaluation}} \\
        \textbf{Method}                        & t-DCF           & \%         & t-DCF           & \% \\ \hline
        Initial                                & .0142 (.0029) & 0.0          & .2315 (.0547) & 0.0          \\ \hline
        \textit{Independent} &&&& \\
        IM-separate                            & .0182 (.0103) & 19.5         & .2248 (.1065) & -7.8       \\
        IM-same                                & .0306 (.0089) & 131.7        & .2629 (.0693) & 13.6        \\ \hline
        \textit{REINFORCE} &&&& \\
        Simple                                 & .0076 (.0023) & -42.0        & \textbf{.1690} (.0359) & \textbf{-26.5}        \\
        Penalize                               & .0074 (.0015) & -45.8        & .1729 (.0318) & -24.5           \\
        Reward                                 & .0075 (.0019) & -43.1        & .1737 (.0332) & -24.2       \\
        t-DCF                                  & \textbf{.0073} (.0011) & \textbf{-45.6}          & .1725 (.0290) & -24.3         
        \end{tabular}
        \tablesqueezevspace
        \caption{{\it Results of tandem training, averaged over three repetitions with standard deviation in the parenthesis. The ``\%" column shows the average, relative change in the t-DCF compared to the initial model in percentages. Note that the development set was used for training.}}
        \label{table:tdcf_results}
    \end{table}

    Table \ref{table:tdcf_results} summarizes the average t-DCFs before and after the tandem optimization with different methods. REINFORCE with different reward functions is able to reach the highest reliable improvement, with \textit{t-DCF} reward function having least variance in its results. Interestingly, training with cross-entropy, the performance degrades in the training set, but in the evaluation set the performance either improves (\textit{IM-same}) or degrades (\textit{IM-separate}).
    
    In the repetition two the t-DCF increased from $0.3087$ to $0.3741$ with \textit{IM-separate} in the evaluation set, while in all REINFORCE runs the t-DCF reduced regardless of the initial parameters. Without this outlier repetition the \textit{IM-separate} obtained an average t-DCF decrease of $22\%$, placing it to the similar levels with REINFORCE methods. For \textit{IM-same} the same treatment still leads to an increased t-DCF of $14\%$. This suggests that REINFORCE-based training offers robust tandem training regardless of what the initial parameters are.
    
    To further analyze how the ASV and CM systems evolved during the training, Fig.~\ref{fig:learning_curves} shows how performance changed over training. The high variance of the performance between consecutive steps is common for REINFORCE \cite{henderson2018deep}. REINFORCE training with all reward functions improves the t-DCF over time, but interestingly ASV performance \textit{decreases} on the evaluation set. This ASV result is a recurring theme over all the training methods and might be due to limited number of speakers in the training set. With the \textit{t-DCF} reward function, t-DCF improves in same degree as with other tested reward functions, but CM EER only reduces by an average $2\%$, while for the other methods this is closer to $10\%$. This suggests that the REINFORCE training with rewards derived from the cost-parameters learns to optimize for a specific operating point, and/or train systems to \say{cooperate together}, without improving the individual systems in their separate tasks. 
    
    Similar to \textit{IM-same} training setup, we hypothesise REINFORCE training allows systems to learn \say{each others'} tasks. To test this hypothesis, we evaluate ASV's performance on CM's task and vice-versa, swapping the ground-truth labels while computing the EER. With all reward functions tested, ASV's EER in the CM task drops by relative $2\%$ and CM's EER in the ASV task drops by relative $10\%$ in the development set. In the evaluation set the EERs drop by average relative $1.5\%$ both ways. The larger decrease of CM's error in ASV on the development set might be due to overfitting, as the training list does not contain many speakers. While not large improvements, this supports our hypothesis that the separate systems learned to do better in each other's tasks, which in turn contributes to the decreased t-DCF. Especially ASV's support in detecting spoof samples helps, as the t-DCF evaluation metric used here is sensitive to CM performance.
    
    \begin{figure}[t]
        \centering
        \subfigure{\includegraphics[width=0.15\textwidth]{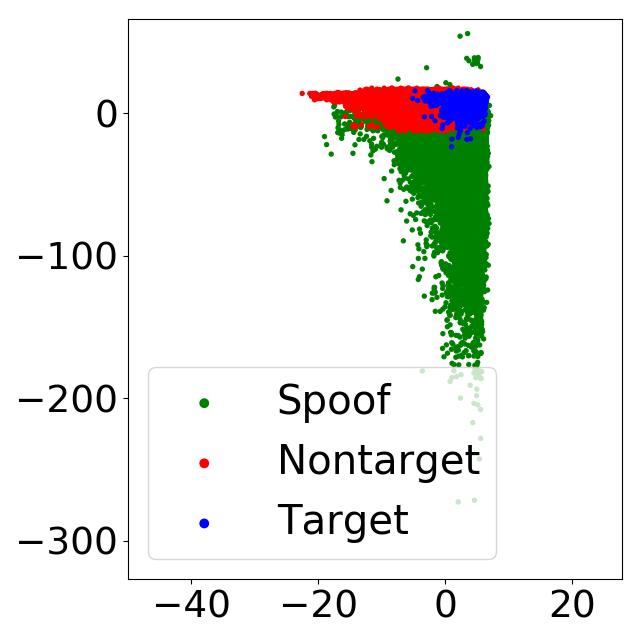}}
        \subfigure{\includegraphics[width=0.15\textwidth]{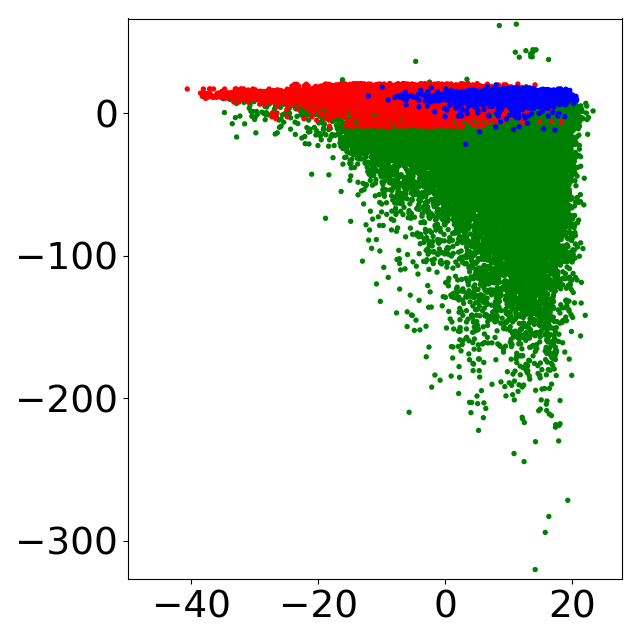}}
        \subfigure{\includegraphics[width=0.15\textwidth]{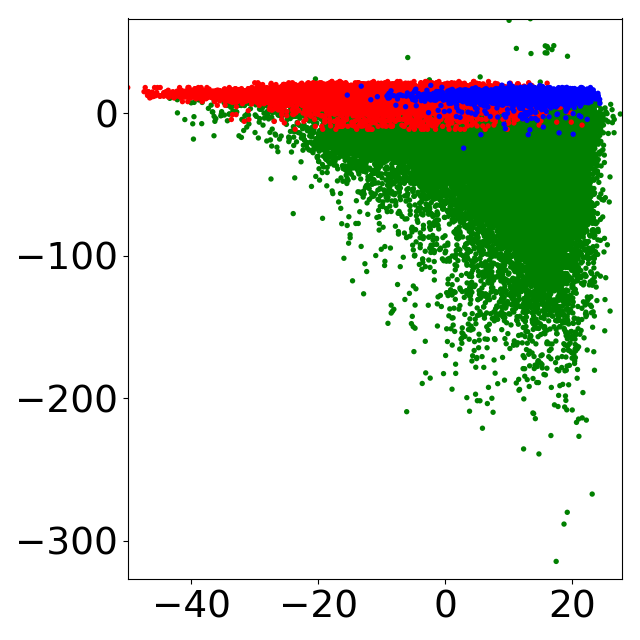}}
        \figsqueezevspace
        \caption{{\it Evolution of ASV and CM scores during REINFORCE \textit{Simple} training in one of the repetitions, from the start to the end of training regimen. ASV scores are on the X-axis and CM scores on Y-axis. Scores are scaled towards higher magnitudes with no clear separation occurring.}}
        \label{fig:score_scatter}
    \end{figure}
    
\begin{figure}[t]
        \centering
        \subfigure[\textit{Simple}]{\includegraphics[width=0.15\textwidth]{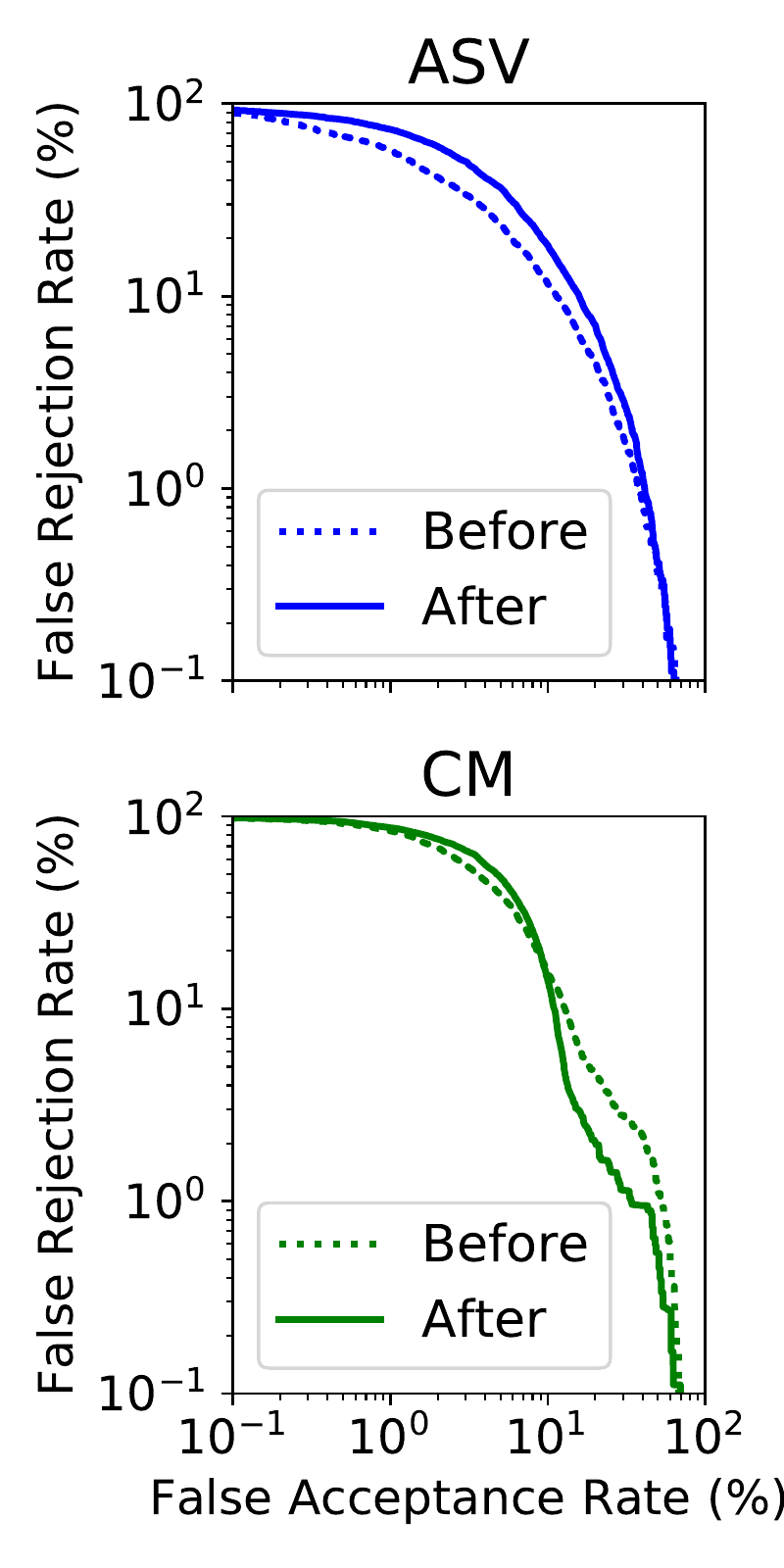}}
        \subfigure[\textit{t-DCF}]{\includegraphics[width=0.15\textwidth]{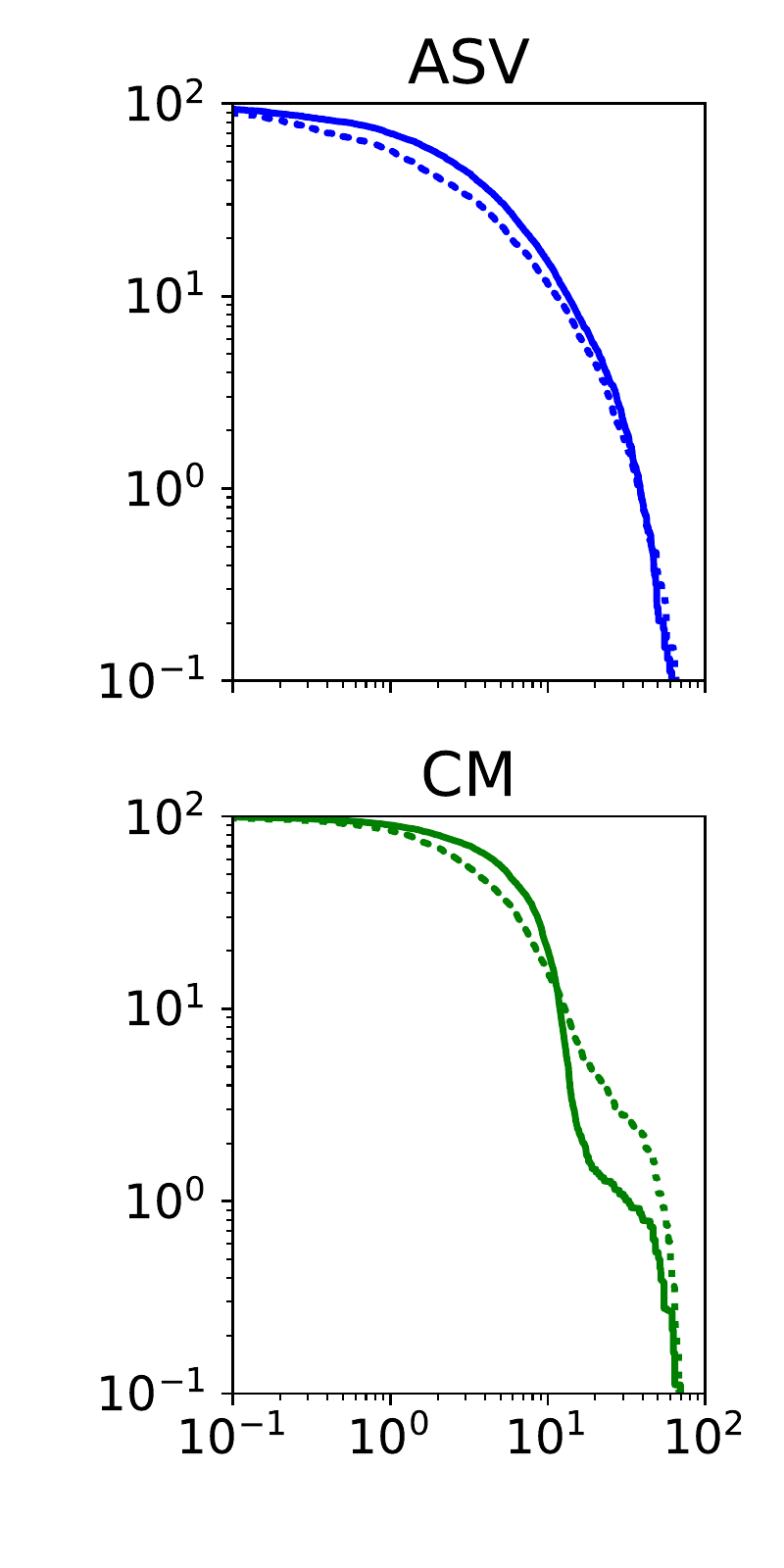}}
        \subfigure[\textit{IM-separate}]{\includegraphics[width=0.15\textwidth]{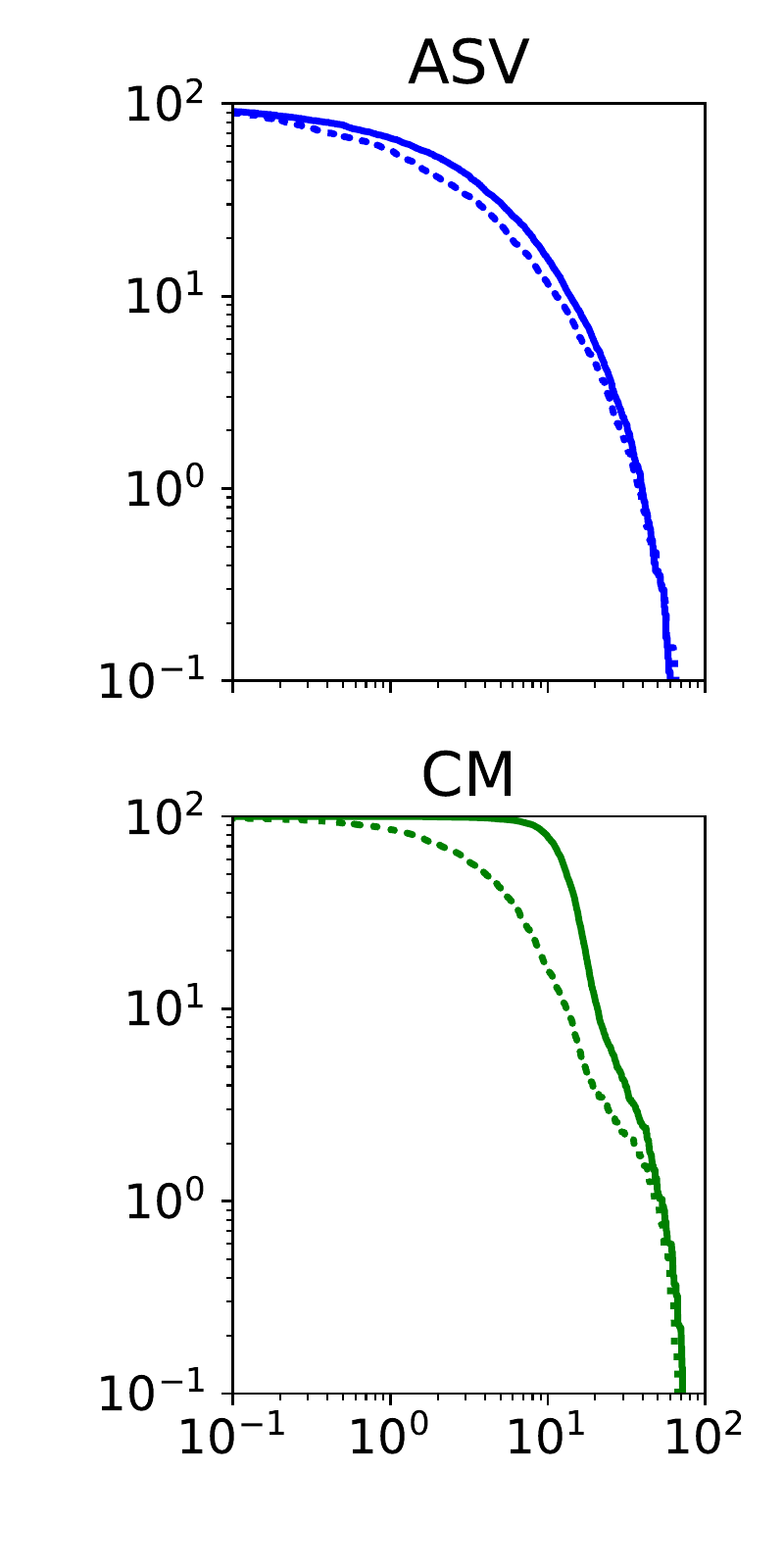}}
        \figsqueezevspace
        \caption{{\it DET curves of ASV and CM systems before and after tandem training, evaluated on the evaluation set and trained on same initial parameters. These figures were chosen to highlight the differences in the results of different training procedures. Where REINFORCE based training is able to improve CM's performance, cross-entropy training only manages to decrease this.}}
        \label{fig:det_changes}
    \end{figure}

    By studying the \textit{detection error tradeoff} (DET) curves of the individual systems before and after tandem optimization (Fig~\ref{fig:det_changes}), we see that CM system's DET curve does not improve (move towards the left-lower corner) overall but improves in specific areas and degrades in others. In contrast, ASV's DET curve either increases or decreases steadily over the range. With \textit{IM-separate} and \textit{IM-same}, the CM FRR increased at CM FARs under $\approx 10\%$ in all repetitions, but at higher FRRs the FAR decreased. Fig.~\ref{fig:det_changes}c shows an example of a situation where \textit{IM-separate} only decreased the overall performance. In REINFORCE based learning, the error-rates decrease around a specific spot with FRR $\approx 1\%$ and/or FAR $\approx 10\%$ (Figures \ref{fig:det_changes}a and \ref{fig:det_changes}b). This also suggests that the training schemes optimize for specific operating points, rather than learning to separate the target/nontarget classes from each other. Intuitively this is understandable, as \textit{e.g.} t-DCF defines the operating by setting how much we allow false-accepts compared to false-rejections with its cost-parameters.
    
    While discussing policy gradient optimization in the Section \ref{sec:method}, we speculated that the REINFORCE algorithm would \say{sharpen} the scores, moving them towards infinity without bounds. Fig.~\ref{fig:score_scatter} demonstrates the evolution of scores predicted by ASV and CM systems during one of the repetitions with REINFORCE \textit{Simple} reward function. The scores indeed spread out without clear separation happening. However, in cross-entropy based methods (\textit{IM-separate} and \textit{IM-same}), the scores first clump up together in the first few updates after which they begin to spread out and separate. We suspect that this is due to cross-entropy moving scores based on how large their error was (loss is higher if error is high), causing initial errors to have a large loss and thus larger changes in the parameters. Meanwhile, REINFORCE-based learning has a fixed step-size per sample, independent of how far-off the prediction was. This suggests REINFORCE-based fine-tuning might be stabler than the standard cross-entropy learning, at the cost of REINFORCE training requiring more training steps.
    
\sectionpresqueeze
\section{Conclusion}
\sectionpostsqueeze
    Speaker verification systems together with a spoofing countermeasures are trained separate from each other, yet they are evaluated as one \say{tandem} system. In this work, we demonstrated how reinforcement learning can be used to optimize such spoof-robust speaker verification tandem system with multiple, individual components. Our results indicate that this REINFORCE method out-performs comparable fine-tuning methods of supervised learning, while retaining the individual systems intact for their respective tasks. Curiously, we found that this tandem optimization led to the separate systems \textit{learning to do each other's tasks}. Further, the optimization led to reduced error-rates at a specific threshold. Overall,  we see a $20\%$ relative decrease in relative t-DCF after tandem optimization, \textit{but with ASV error-rates increasing by relative $10\%$}. This suggests the systems learned to cooperate for better tandem performance, rather than just improving in their separate tasks. 
    
    This is but an initial study on using such methods on tandem systems in speaker verification and related topics. For example, we did not study the effect of dataset bias on the REINFORCE training, and how it would change the operating point we are optimizing for. We also used a single set of models and datasets; future work should be replicated on alternative ASV and CM models and datasets. Specifically, what would happen if the pre-trained systems already had near-perfect error-rates? Would such tandem-optimization still improve the system as a whole? 
    
    REINFORCE training could also be studied further with \textit{e.g.} alternative sampling with differentiable distribution known as Gumbell-softmax \cite{jang2017categorical}, which results in less variance than REINFORCE. One could also create a \say{soft t-DCF} (\textit{i.e.} differentiable t-DCF) which could be optimized directly without REINFORCE, something we hinted at in the introduction. While many questions remain, we are motivated to study this topic further in the light of the positive results achieved here.
    
\bibliographystyle{IEEEbib}
\bibliography{refs.bib}

\end{document}